\newcommand\papertitle{\Planck~intermediate results. XII: Diffuse Galactic components in the Gould Belt System}

\documentclass[longauth,twocolumn,traditabstract]{aa} % for a paper on 1 column  
\usepackage{graphicx}
\usepackage{txfonts}
\usepackage[hyphens]{url}
\usepackage{natbib}
\bibpunct{(}{)}{;}{a}{}{,}  % to follow the A&A style
\def\setsymbol#1#2{\expandafter\def\csname #1\endcsname{#2}}
\def\getsymbol#1{\csname #1\endcsname}

\def\Planck{\textit{Planck}}

\newbox\tablebox    \newdimen\tablewidth
\def\leaderfil{\leaders\hbox to 5pt{\hss.\hss}\hfil}
\def\endPlancktablewide{\tablewidth=\textwidth 
    $$\hss\copy\tablebox\hss$$
    \vskip-\lastskip\vskip -2pt}
\def\tablenote#1 #2\par{\begingroup \parindent=0.8em
    \abovedisplayshortskip=0pt\belowdisplayshortskip=0pt
    \noindent
    $$\hss\vbox{\hsize\tablewidth \hangindent=\parindent \hangafter=1 \noindent
    \hbox to \parindent{$^#1$\hss}\strut#2\strut\par}\hss$$
    \endgroup}
\def\doubleline{\vskip 3pt\hrule \vskip 1.5pt \hrule \vskip 5pt}

\def\L2{\ifmmode L_2\else $L_2$\fi}

\def\DeltaT{\ifmmode \Delta T\else $\Delta T$\fi}
\def\deltat{\ifmmode \Delta t\else $\Delta t$\fi}
\def\fknee{\ifmmode f_{\rm knee}\else $f_{\rm knee}$\fi}
\def\Fmax{\ifmmode F_{\rm max}\else $F_{\rm max}$\fi}
\def\solar{\ifmmode{\rm M}_{\mathord\odot}\else${\rm M}_{\mathord\odot}$\fi}
\def\Msolar{\ifmmode{\rm M}_{\mathord\odot}\else${\rm M}_{\mathord\odot}$\fi}
\def\Lsolar{\ifmmode{\rm L}_{\mathord\odot}\else${\rm L}_{\mathord\odot}$\fi}

\def\inv{\ifmmode^{-1}\else$^{-1}$\fi}
\def\mo{\ifmmode^{-1}\else$^{-1}$\fi}
\def\sup#1{\ifmmode ^{\rm #1}\else $^{\rm #1}$\fi}
\def\expo#1{\ifmmode \times 10^{#1}\else $\times 10^{#1}$\fi}
\def\,{\thinspace}
\def\lsim{\mathrel{\raise .4ex\hbox{\rlap{$<$}\lower 1.2ex\hbox{$\sim$}}}}
\def\gsim{\mathrel{\raise .4ex\hbox{\rlap{$>$}\lower 1.2ex\hbox{$\sim$}}}}

\def\simprop{\mathrel{\raise .4ex\hbox{\rlap{$\propto$}\lower 1.2ex\hbox{$\sim$}}}}
\def\deg{\ifmmode^\circ\else$^\circ$\fi}
\def\pdeg{\ifmmode $\setbox0=\hbox{$^{\circ}$}\rlap{\hskip.11\wd0 .}$^{\circ}
          \else \setbox0=\hbox{$^{\circ}$}\rlap{\hskip.11\wd0 .}$^{\circ}$\fi}
\def\arcs{\ifmmode {^{\scriptstyle\prime\prime}}
          \else $^{\scriptstyle\prime\prime}$\fi}
\def\arcm{\ifmmode {^{\scriptstyle\prime}}
          \else $^{\scriptstyle\prime}$\fi}
\newdimen\sa  \newdimen\sb
\def\parcs{\sa=.07em \sb=.03em
     \ifmmode \hbox{\rlap{.}}^{\scriptstyle\prime\kern -\sb\prime}\hbox{\kern -\sa}
     \else \rlap{.}$^{\scriptstyle\prime\kern -\sb\prime}$\kern -\sa\fi}
\def\parcm{\sa=.08em \sb=.03em
     \ifmmode \hbox{\rlap{.}\kern\sa}^{\scriptstyle\prime}\hbox{\kern-\sb}
     \else \rlap{.}\kern\sa$^{\scriptstyle\prime}$\kern-\sb\fi}
\def\ra[#1 #2 #3.#4]{#1\sup{h}#2\sup{m}#3\sup{s}\llap.#4}
\def\dec[#1 #2 #3.#4]{#1\deg#2\arcm#3\arcs\llap.#4}
\def\deco[#1 #2 #3]{#1\deg#2\arcm#3\arcs}
\def\rra[#1 #2]{#1\sup{h}#2\sup{m}}

\def\dots{\relax\ifmmode \ldots\else $\ldots$\fi}
\def\WHzsr{\ifmmode $W\,Hz\mo\,sr\mo$\else W\,Hz\mo\,sr\mo\fi}
\def\mHz{\ifmmode $\,mHz$\else \,mHz\fi}
\def\GHz{\ifmmode $\,GHz$\else \,GHz\fi}
\def\mKs{\ifmmode $\,mK\,s$^{1/2}\else \,mK\,s$^{1/2}$\fi}
\def\muKs{\ifmmode \,\mu$K\,s$^{1/2}\else \,$\mu$K\,s$^{1/2}$\fi}
\def\muKRJs{\ifmmode \,\mu$K$_{\rm RJ}$\,s$^{1/2}\else \,$\mu$K$_{\rm RJ}$\,s$^{1/2}$\fi}
\def\muKHz{\ifmmode \,\mu$K\,Hz$^{-1/2}\else \,$\mu$K\,Hz$^{-1/2}$\fi}
\def\MJysr{\ifmmode \,$MJy\,sr\mo$\else \,MJy\,sr\mo\fi}
\def\MJysrmK{\ifmmode \,$MJy\,sr\mo$\,mK$_{\rm CMB}\mo\else \,MJy\,sr\mo\,mK$_{\rm CMB}\mo$\fi}
\def\microns{\ifmmode \,\mu$m$\else \,$\mu$m\fi}

\def\muK{\ifmmode \,\mu$K$\else \,$\mu$\hbox{K}\fi}
\def\microK{\ifmmode \,\mu$K$\else \,$\mu$\hbox{K}\fi}
\def\muW{\ifmmode \,\mu$W$\else \,$\mu$\hbox{W}\fi}
\def\kms{\ifmmode $\,km\,s$^{-1}\else \,km\,s$^{-1}$\fi}
\def\kmsMpc{\ifmmode $\,\kms\,Mpc\mo$\else \,\kms\,Mpc\mo\fi}
\setsymbol{LFI:center:frequency:70GHz:units}{70.3\,GHz}
\setsymbol{LFI:center:frequency:44GHz:units}{44.1\,GHz}
\setsymbol{LFI:center:frequency:30GHz:units}{28.5\,GHz}

\setsymbol{LFI:center:frequency:70GHz}{70.3}
\setsymbol{LFI:center:frequency:44GHz}{44.1}
\setsymbol{LFI:center:frequency:30GHz}{28.5}

\setsymbol{LFI:center:frequency:LFI18:Rad:M:units}{71.7\GHz}
\setsymbol{LFI:center:frequency:LFI19:Rad:M:units}{67.5\GHz}
\setsymbol{LFI:center:frequency:LFI20:Rad:M:units}{69.2\GHz}
\setsymbol{LFI:center:frequency:LFI21:Rad:M:units}{70.4\GHz}
\setsymbol{LFI:center:frequency:LFI22:Rad:M:units}{71.5\GHz}
\setsymbol{LFI:center:frequency:LFI23:Rad:M:units}{70.8\GHz}
\setsymbol{LFI:center:frequency:LFI24:Rad:M:units}{44.4\GHz}
\setsymbol{LFI:center:frequency:LFI25:Rad:M:units}{44.0\GHz}
\setsymbol{LFI:center:frequency:LFI26:Rad:M:units}{43.9\GHz}
\setsymbol{LFI:center:frequency:LFI27:Rad:M:units}{28.3\GHz}
\setsymbol{LFI:center:frequency:LFI28:Rad:M:units}{28.8\GHz}
\setsymbol{LFI:center:frequency:LFI18:Rad:S:units}{70.1\GHz}
\setsymbol{LFI:center:frequency:LFI19:Rad:S:units}{69.6\GHz}
\setsymbol{LFI:center:frequency:LFI20:Rad:S:units}{69.5\GHz}
\setsymbol{LFI:center:frequency:LFI21:Rad:S:units}{69.5\GHz}
\setsymbol{LFI:center:frequency:LFI22:Rad:S:units}{72.8\GHz}
\setsymbol{LFI:center:frequency:LFI23:Rad:S:units}{71.3\GHz}
\setsymbol{LFI:center:frequency:LFI24:Rad:S:units}{44.1\GHz}
\setsymbol{LFI:center:frequency:LFI25:Rad:S:units}{44.1\GHz}
\setsymbol{LFI:center:frequency:LFI26:Rad:S:units}{44.1\GHz}
\setsymbol{LFI:center:frequency:LFI27:Rad:S:units}{28.5\GHz}
\setsymbol{LFI:center:frequency:LFI28:Rad:S:units}{28.2\GHz}

\setsymbol{LFI:center:frequency:LFI18:Rad:M}{71.7}
\setsymbol{LFI:center:frequency:LFI19:Rad:M}{67.5}
\setsymbol{LFI:center:frequency:LFI20:Rad:M}{69.2}
\setsymbol{LFI:center:frequency:LFI21:Rad:M}{70.4}
\setsymbol{LFI:center:frequency:LFI22:Rad:M}{71.5}
\setsymbol{LFI:center:frequency:LFI23:Rad:M}{70.8}
\setsymbol{LFI:center:frequency:LFI24:Rad:M}{44.4}
\setsymbol{LFI:center:frequency:LFI25:Rad:M}{44.0}
\setsymbol{LFI:center:frequency:LFI26:Rad:M}{43.9}
\setsymbol{LFI:center:frequency:LFI27:Rad:M}{28.3}
\setsymbol{LFI:center:frequency:LFI28:Rad:M}{28.8}
\setsymbol{LFI:center:frequency:LFI18:Rad:S}{70.1}
\setsymbol{LFI:center:frequency:LFI19:Rad:S}{69.6}
\setsymbol{LFI:center:frequency:LFI20:Rad:S}{69.5}
\setsymbol{LFI:center:frequency:LFI21:Rad:S}{69.5}
\setsymbol{LFI:center:frequency:LFI22:Rad:S}{72.8}
\setsymbol{LFI:center:frequency:LFI23:Rad:S}{71.3}
\setsymbol{LFI:center:frequency:LFI24:Rad:S}{44.1}
\setsymbol{LFI:center:frequency:LFI25:Rad:S}{44.1}
\setsymbol{LFI:center:frequency:LFI26:Rad:S}{44.1}
\setsymbol{LFI:center:frequency:LFI27:Rad:S}{28.5}
\setsymbol{LFI:center:frequency:LFI28:Rad:S}{28.2}

\setsymbol{LFI:white:noise:sensitivity:70GHz:units}{134.7\muKs}
\setsymbol{LFI:white:noise:sensitivity:44GHz:units}{164.7\muKs}
\setsymbol{LFI:white:noise:sensitivity:30GHz:units}{143.4\muKs}

\setsymbol{LFI:white:noise:sensitivity:70GHz}{134.7}
\setsymbol{LFI:white:noise:sensitivity:44GHz}{164.7}
\setsymbol{LFI:white:noise:sensitivity:30GHz}{143.4}

\setsymbol{LFI:white:noise:sensitivity:LFI18:Rad:M:units}{512.0\muKs}
\setsymbol{LFI:white:noise:sensitivity:LFI19:Rad:M:units}{581.4\muKs}
\setsymbol{LFI:white:noise:sensitivity:LFI20:Rad:M:units}{590.8\muKs}
\setsymbol{LFI:white:noise:sensitivity:LFI21:Rad:M:units}{455.2\muKs}
\setsymbol{LFI:white:noise:sensitivity:LFI22:Rad:M:units}{492.0\muKs}
\setsymbol{LFI:white:noise:sensitivity:LFI23:Rad:M:units}{507.7\muKs}
\setsymbol{LFI:white:noise:sensitivity:LFI24:Rad:M:units}{462.2\muKs}
\setsymbol{LFI:white:noise:sensitivity:LFI25:Rad:M:units}{413.6\muKs}
\setsymbol{LFI:white:noise:sensitivity:LFI26:Rad:M:units}{478.6\muKs}
\setsymbol{LFI:white:noise:sensitivity:LFI27:Rad:M:units}{277.7\muKs}
\setsymbol{LFI:white:noise:sensitivity:LFI28:Rad:M:units}{312.3\muKs}
\setsymbol{LFI:white:noise:sensitivity:LFI18:Rad:S:units}{465.7\muKs}
\setsymbol{LFI:white:noise:sensitivity:LFI19:Rad:S:units}{555.6\muKs}
\setsymbol{LFI:white:noise:sensitivity:LFI20:Rad:S:units}{623.2\muKs}
\setsymbol{LFI:white:noise:sensitivity:LFI21:Rad:S:units}{564.1\muKs}
\setsymbol{LFI:white:noise:sensitivity:LFI22:Rad:S:units}{534.4\muKs}
\setsymbol{LFI:white:noise:sensitivity:LFI23:Rad:S:units}{542.4\muKs}
\setsymbol{LFI:white:noise:sensitivity:LFI24:Rad:S:units}{399.2\muKs}
\setsymbol{LFI:white:noise:sensitivity:LFI25:Rad:S:units}{392.6\muKs}
\setsymbol{LFI:white:noise:sensitivity:LFI26:Rad:S:units}{418.6\muKs}
\setsymbol{LFI:white:noise:sensitivity:LFI27:Rad:S:units}{302.9\muKs}
\setsymbol{LFI:white:noise:sensitivity:LFI28:Rad:S:units}{285.3\muKs}

\setsymbol{LFI:white:noise:sensitivity:LFI18:Rad:M}{512.0}
\setsymbol{LFI:white:noise:sensitivity:LFI19:Rad:M}{581.4}
\setsymbol{LFI:white:noise:sensitivity:LFI20:Rad:M}{590.8}
\setsymbol{LFI:white:noise:sensitivity:LFI21:Rad:M}{455.2}
\setsymbol{LFI:white:noise:sensitivity:LFI22:Rad:M}{492.0}
\setsymbol{LFI:white:noise:sensitivity:LFI23:Rad:M}{507.7}
\setsymbol{LFI:white:noise:sensitivity:LFI24:Rad:M}{462.2}
\setsymbol{LFI:white:noise:sensitivity:LFI25:Rad:M}{413.6}
\setsymbol{LFI:white:noise:sensitivity:LFI26:Rad:M}{478.6}
\setsymbol{LFI:white:noise:sensitivity:LFI27:Rad:M}{277.7}
\setsymbol{LFI:white:noise:sensitivity:LFI28:Rad:M}{312.3}
\setsymbol{LFI:white:noise:sensitivity:LFI18:Rad:S}{465.7}
\setsymbol{LFI:white:noise:sensitivity:LFI19:Rad:S}{555.6}
\setsymbol{LFI:white:noise:sensitivity:LFI20:Rad:S}{623.2}
\setsymbol{LFI:white:noise:sensitivity:LFI21:Rad:S}{564.1}
\setsymbol{LFI:white:noise:sensitivity:LFI22:Rad:S}{534.4}
\setsymbol{LFI:white:noise:sensitivity:LFI23:Rad:S}{542.4}
\setsymbol{LFI:white:noise:sensitivity:LFI24:Rad:S}{399.2}
\setsymbol{LFI:white:noise:sensitivity:LFI25:Rad:S}{392.6}
\setsymbol{LFI:white:noise:sensitivity:LFI26:Rad:S}{418.6}
\setsymbol{LFI:white:noise:sensitivity:LFI27:Rad:S}{302.9}
\setsymbol{LFI:white:noise:sensitivity:LFI28:Rad:S}{285.3}

\setsymbol{LFI:knee:frequency:70GHz:units}{29.5\mHz}
\setsymbol{LFI:knee:frequency:44GHz:units}{56.2\mHz}
\setsymbol{LFI:knee:frequency:30GHz:units}{113.7\mHz}

\setsymbol{LFI:knee:frequency:70GHz}{29.5}
\setsymbol{LFI:knee:frequency:44GHz}{56.2}
\setsymbol{LFI:knee:frequency:30GHz}{113.7}

\setsymbol{LFI:knee:frequency:LFI18:Rad:M:units}{16.3\mHz}
\setsymbol{LFI:knee:frequency:LFI19:Rad:M:units}{15.1\mHz}
\setsymbol{LFI:knee:frequency:LFI20:Rad:M:units}{18.7\mHz}
\setsymbol{LFI:knee:frequency:LFI21:Rad:M:units}{37.2\mHz}
\setsymbol{LFI:knee:frequency:LFI22:Rad:M:units}{12.7\mHz}
\setsymbol{LFI:knee:frequency:LFI23:Rad:M:units}{34.6\mHz}
\setsymbol{LFI:knee:frequency:LFI24:Rad:M:units}{46.2\mHz}
\setsymbol{LFI:knee:frequency:LFI25:Rad:M:units}{24.9\mHz}
\setsymbol{LFI:knee:frequency:LFI26:Rad:M:units}{67.6\mHz}
\setsymbol{LFI:knee:frequency:LFI27:Rad:M:units}{187.4\mHz}
\setsymbol{LFI:knee:frequency:LFI28:Rad:M:units}{122.2\mHz}
\setsymbol{LFI:knee:frequency:LFI18:Rad:S:units}{17.7\mHz}
\setsymbol{LFI:knee:frequency:LFI19:Rad:S:units}{22.0\mHz}
\setsymbol{LFI:knee:frequency:LFI20:Rad:S:units}{8.7\mHz}
\setsymbol{LFI:knee:frequency:LFI21:Rad:S:units}{25.9\mHz}
\setsymbol{LFI:knee:frequency:LFI22:Rad:S:units}{15.8\mHz}
\setsymbol{LFI:knee:frequency:LFI23:Rad:S:units}{129.8\mHz}
\setsymbol{LFI:knee:frequency:LFI24:Rad:S:units}{100.9\mHz}
\setsymbol{LFI:knee:frequency:LFI25:Rad:S:units}{38.9\mHz}
\setsymbol{LFI:knee:frequency:LFI26:Rad:S:units}{58.9\mHz}
\setsymbol{LFI:knee:frequency:LFI27:Rad:S:units}{104.4\mHz}
\setsymbol{LFI:knee:frequency:LFI28:Rad:S:units}{40.7\mHz}

\setsymbol{LFI:knee:frequency:LFI18:Rad:M}{16.3}
\setsymbol{LFI:knee:frequency:LFI19:Rad:M}{15.1}
\setsymbol{LFI:knee:frequency:LFI20:Rad:M}{18.7}
\setsymbol{LFI:knee:frequency:LFI21:Rad:M}{37.2}
\setsymbol{LFI:knee:frequency:LFI22:Rad:M}{12.7}
\setsymbol{LFI:knee:frequency:LFI23:Rad:M}{34.6}
\setsymbol{LFI:knee:frequency:LFI24:Rad:M}{46.2}
\setsymbol{LFI:knee:frequency:LFI25:Rad:M}{24.9}
\setsymbol{LFI:knee:frequency:LFI26:Rad:M}{67.6}
\setsymbol{LFI:knee:frequency:LFI27:Rad:M}{187.4}
\setsymbol{LFI:knee:frequency:LFI28:Rad:M}{122.2}
\setsymbol{LFI:knee:frequency:LFI18:Rad:S}{17.7}
\setsymbol{LFI:knee:frequency:LFI19:Rad:S}{22.0}
\setsymbol{LFI:knee:frequency:LFI20:Rad:S}{8.7}
\setsymbol{LFI:knee:frequency:LFI21:Rad:S}{25.9}
\setsymbol{LFI:knee:frequency:LFI22:Rad:S}{15.8}
\setsymbol{LFI:knee:frequency:LFI23:Rad:S}{129.8}
\setsymbol{LFI:knee:frequency:LFI24:Rad:S}{100.9}
\setsymbol{LFI:knee:frequency:LFI25:Rad:S}{38.9}
\setsymbol{LFI:knee:frequency:LFI26:Rad:S}{58.9}
\setsymbol{LFI:knee:frequency:LFI27:Rad:S}{104.4}
\setsymbol{LFI:knee:frequency:LFI28:Rad:S}{40.7}

\setsymbol{LFI:slope:70GHz:units}{$-1.03$\mHz}
\setsymbol{LFI:slope:44GHz:units}{$-0.89$\mHz}
\setsymbol{LFI:slope:30GHz:units}{$-0.87$\mHz}

\setsymbol{LFI:slope:70GHz}{$-1.03$}
\setsymbol{LFI:slope:44GHz}{$-0.89$}
\setsymbol{LFI:slope:30GHz}{$-0.87$}

\setsymbol{LFI:slope:LFI18:Rad:M:units}{$-1.04$\mHz}
\setsymbol{LFI:slope:LFI19:Rad:M:units}{$-1.09$\mHz}
\setsymbol{LFI:slope:LFI20:Rad:M:units}{$-0.69$\mHz}
\setsymbol{LFI:slope:LFI21:Rad:M:units}{$-1.56$\mHz}
\setsymbol{LFI:slope:LFI22:Rad:M:units}{$-1.01$\mHz}
\setsymbol{LFI:slope:LFI23:Rad:M:units}{$-0.96$\mHz}
\setsymbol{LFI:slope:LFI24:Rad:M:units}{$-0.83$\mHz}
\setsymbol{LFI:slope:LFI25:Rad:M:units}{$-0.91$\mHz}
\setsymbol{LFI:slope:LFI26:Rad:M:units}{$-0.95$\mHz}
\setsymbol{LFI:slope:LFI27:Rad:M:units}{$-0.87$\mHz}
\setsymbol{LFI:slope:LFI28:Rad:M:units}{$-0.88$\mHz}
\setsymbol{LFI:slope:LFI18:Rad:S:units}{$-1.15$\mHz}
\setsymbol{LFI:slope:LFI19:Rad:S:units}{$-1.00$\mHz}
\setsymbol{LFI:slope:LFI20:Rad:S:units}{$-0.95$\mHz}
\setsymbol{LFI:slope:LFI21:Rad:S:units}{$-0.92$\mHz}
\setsymbol{LFI:slope:LFI22:Rad:S:units}{$-1.01$\mHz}
\setsymbol{LFI:slope:LFI23:Rad:S:units}{$-0.95$\mHz}
\setsymbol{LFI:slope:LFI24:Rad:S:units}{$-0.73$\mHz}
\setsymbol{LFI:slope:LFI25:Rad:S:units}{$-1.16$\mHz}
\setsymbol{LFI:slope:LFI26:Rad:S:units}{$-0.79$\mHz}
\setsymbol{LFI:slope:LFI27:Rad:S:units}{$-0.82$\mHz}
\setsymbol{LFI:slope:LFI28:Rad:S:units}{$-0.91$\mHz}

\setsymbol{LFI:slope:LFI18:Rad:M}{$-1.04$}
\setsymbol{LFI:slope:LFI19:Rad:M}{$-1.09$}
\setsymbol{LFI:slope:LFI20:Rad:M}{$-0.69$}
\setsymbol{LFI:slope:LFI21:Rad:M}{$-1.56$}
\setsymbol{LFI:slope:LFI22:Rad:M}{$-1.01$}
\setsymbol{LFI:slope:LFI23:Rad:M}{$-0.96$}
\setsymbol{LFI:slope:LFI24:Rad:M}{$-0.83$}
\setsymbol{LFI:slope:LFI25:Rad:M}{$-0.91$}
\setsymbol{LFI:slope:LFI26:Rad:M}{$-0.95$}
\setsymbol{LFI:slope:LFI27:Rad:M}{$-0.87$}
\setsymbol{LFI:slope:LFI28:Rad:M}{$-0.88$}
\setsymbol{LFI:slope:LFI18:Rad:S}{$-1.15$}
\setsymbol{LFI:slope:LFI19:Rad:S}{$-1.00$}
\setsymbol{LFI:slope:LFI20:Rad:S}{$-0.95$}
\setsymbol{LFI:slope:LFI21:Rad:S}{$-0.92$}
\setsymbol{LFI:slope:LFI22:Rad:S}{$-1.01$}
\setsymbol{LFI:slope:LFI23:Rad:S}{$-0.95$}
\setsymbol{LFI:slope:LFI24:Rad:S}{$-0.73$}
\setsymbol{LFI:slope:LFI25:Rad:S}{$-1.16$}
\setsymbol{LFI:slope:LFI26:Rad:S}{$-0.79$}
\setsymbol{LFI:slope:LFI27:Rad:S}{$-0.82$}
\setsymbol{LFI:slope:LFI28:Rad:S}{$-0.91$}

\setsymbol{LFI:FWHM:70GHz:units}{13\parcm01}
\setsymbol{LFI:FWHM:44GHz:units}{27\parcm92}
\setsymbol{LFI:FWHM:30GHz:units}{32\parcm65}

\setsymbol{LFI:FWHM:70GHz}{13.01}
\setsymbol{LFI:FWHM:44GHz}{27.92}
\setsymbol{LFI:FWHM:30GHz}{32.65}

\setsymbol{LFI:FWHM:LFI18:units}{13\parcm39}
\setsymbol{LFI:FWHM:LFI19:units}{13\parcm01}
\setsymbol{LFI:FWHM:LFI20:units}{12\parcm75}
\setsymbol{LFI:FWHM:LFI21:units}{12\parcm74}
\setsymbol{LFI:FWHM:LFI22:units}{12\parcm87}
\setsymbol{LFI:FWHM:LFI23:units}{13\parcm27}
\setsymbol{LFI:FWHM:LFI24:units}{22\parcm98}
\setsymbol{LFI:FWHM:LFI25:units}{30\parcm46}
\setsymbol{LFI:FWHM:LFI26:units}{30\parcm31}
\setsymbol{LFI:FWHM:LFI27:units}{32\parcm65}
\setsymbol{LFI:FWHM:LFI28:units}{32\parcm66}

\setsymbol{LFI:FWHM:LFI18}{13.39}
\setsymbol{LFI:FWHM:LFI19}{13.01}
\setsymbol{LFI:FWHM:LFI20}{12.75}
\setsymbol{LFI:FWHM:LFI21}{12.74}
\setsymbol{LFI:FWHM:LFI22}{12.87}
\setsymbol{LFI:FWHM:LFI23}{13.27}
\setsymbol{LFI:FWHM:LFI24}{22.98}
\setsymbol{LFI:FWHM:LFI25}{30.46}
\setsymbol{LFI:FWHM:LFI26}{30.31}
\setsymbol{LFI:FWHM:LFI27}{32.65}
\setsymbol{LFI:FWHM:LFI28}{32.66}

\setsymbol{LFI:FWHM:uncertainty:LFI18:units}{0.170\arcm}
\setsymbol{LFI:FWHM:uncertainty:LFI19:units}{0.174\arcm}
\setsymbol{LFI:FWHM:uncertainty:LFI20:units}{0.170\arcm}
\setsymbol{LFI:FWHM:uncertainty:LFI21:units}{0.156\arcm}
\setsymbol{LFI:FWHM:uncertainty:LFI22:units}{0.164\arcm}
\setsymbol{LFI:FWHM:uncertainty:LFI23:units}{0.171\arcm}
\setsymbol{LFI:FWHM:uncertainty:LFI24:units}{0.652\arcm}
\setsymbol{LFI:FWHM:uncertainty:LFI25:units}{1.075\arcm}
\setsymbol{LFI:FWHM:uncertainty:LFI26:units}{1.131\arcm}
\setsymbol{LFI:FWHM:uncertainty:LFI27:units}{1.266\arcm}
\setsymbol{LFI:FWHM:uncertainty:LFI28:units}{1.287\arcm}

\setsymbol{LFI:FWHM:uncertainty:LFI18}{0.170}
\setsymbol{LFI:FWHM:uncertainty:LFI19}{0.174}
\setsymbol{LFI:FWHM:uncertainty:LFI20}{0.170}
\setsymbol{LFI:FWHM:uncertainty:LFI21}{0.156}
\setsymbol{LFI:FWHM:uncertainty:LFI22}{0.164}
\setsymbol{LFI:FWHM:uncertainty:LFI23}{0.171}
\setsymbol{LFI:FWHM:uncertainty:LFI24}{0.652}
\setsymbol{LFI:FWHM:uncertainty:LFI25}{1.075}
\setsymbol{LFI:FWHM:uncertainty:LFI26}{1.131}
\setsymbol{LFI:FWHM:uncertainty:LFI27}{1.266}
\setsymbol{LFI:FWHM:uncertainty:LFI28}{1.287}

\setsymbol{HFI:center:frequency:100GHz:units}{100\,GHz}
\setsymbol{HFI:center:frequency:143GHz:units}{143\,GHz}
\setsymbol{HFI:center:frequency:217GHz:units}{217\,GHz}
\setsymbol{HFI:center:frequency:353GHz:units}{353\,GHz}
\setsymbol{HFI:center:frequency:545GHz:units}{545\,GHz}
\setsymbol{HFI:center:frequency:857GHz:units}{857\,GHz}

\setsymbol{HFI:center:frequency:100GHz}{100}
\setsymbol{HFI:center:frequency:143GHz}{143}
\setsymbol{HFI:center:frequency:217GHz}{217}
\setsymbol{HFI:center:frequency:353GHz}{353}
\setsymbol{HFI:center:frequency:545GHz}{545}
\setsymbol{HFI:center:frequency:857GHz}{857}

\setsymbol{HFI:Ndetectors:100GHz}{8}
\setsymbol{HFI:Ndetectors:143GHz}{11}
\setsymbol{HFI:Ndetectors:217GHz}{12}
\setsymbol{HFI:Ndetectors:353GHz}{12}
\setsymbol{HFI:Ndetectors:545GHz}{3}
\setsymbol{HFI:Ndetectors:857GHz}{4}

\setsymbol{HFI:FWHM:Maps:100GHz:units}{9\parcm88}
\setsymbol{HFI:FWHM:Maps:143GHz:units}{7\parcm18}
\setsymbol{HFI:FWHM:Maps:217GHz:units}{4\parcm87}
\setsymbol{HFI:FWHM:Maps:353GHz:units}{4\parcm65}
\setsymbol{HFI:FWHM:Maps:545GHz:units}{4\parcm72}
\setsymbol{HFI:FWHM:Maps:857GHz:units}{4\parcm39}
\setsymbol{HFI:FWHM:Maps:100GHz}{9.88}
\setsymbol{HFI:FWHM:Maps:143GHz}{7.18}
\setsymbol{HFI:FWHM:Maps:217GHz}{4.87}
\setsymbol{HFI:FWHM:Maps:353GHz}{4.65}
\setsymbol{HFI:FWHM:Maps:545GHz}{4.72}
\setsymbol{HFI:FWHM:Maps:857GHz}{4.39}

\setsymbol{HFI:beam:ellipticity:Maps:100GHz}{1.15}
\setsymbol{HFI:beam:ellipticity:Maps:143GHz}{1.01}
\setsymbol{HFI:beam:ellipticity:Maps:217GHz}{1.06}
\setsymbol{HFI:beam:ellipticity:Maps:353GHz}{1.05}
\setsymbol{HFI:beam:ellipticity:Maps:545GHz}{1.14}
\setsymbol{HFI:beam:ellipticity:Maps:857GHz}{1.19}

\setsymbol{HFI:FWHM:Mars:100GHz:units}{9\parcm37}
\setsymbol{HFI:FWHM:Mars:143GHz:units}{7\parcm04}
\setsymbol{HFI:FWHM:Mars:217GHz:units}{4\parcm68}
\setsymbol{HFI:FWHM:Mars:353GHz:units}{4\parcm43}
\setsymbol{HFI:FWHM:Mars:545GHz:units}{3\parcm80}
\setsymbol{HFI:FWHM:Mars:857GHz:units}{3\parcm67}

\setsymbol{HFI:FWHM:Mars:100GHz}{9.37}
\setsymbol{HFI:FWHM:Mars:143GHz}{7.04}
\setsymbol{HFI:FWHM:Mars:217GHz}{4.68}
\setsymbol{HFI:FWHM:Mars:353GHz}{4.43}
\setsymbol{HFI:FWHM:Mars:545GHz}{3.80}
\setsymbol{HFI:FWHM:Mars:857GHz}{3.67}

\setsymbol{HFI:beam:ellipticity:Mars:100GHz}{1.18}
\setsymbol{HFI:beam:ellipticity:Mars:143GHz}{1.03}
\setsymbol{HFI:beam:ellipticity:Mars:217GHz}{1.14}
\setsymbol{HFI:beam:ellipticity:Mars:353GHz}{1.09}
\setsymbol{HFI:beam:ellipticity:Mars:545GHz}{1.25}
\setsymbol{HFI:beam:ellipticity:Mars:857GHz}{1.03}

\setsymbol{HFI:CMB:relative:calibration:100GHz}{$\lsim 1\%$}
\setsymbol{HFI:CMB:relative:calibration:143GHz}{$\lsim 1\%$}
\setsymbol{HFI:CMB:relative:calibration:217GHz}{$\lsim 1\%$}
\setsymbol{HFI:CMB:relative:calibration:353GHz}{$\lsim 1\%$}
\setsymbol{HFI:CMB:relative:calibration:545GHz}{}
\setsymbol{HFI:CMB:relative:calibration:857GHz}{}

\setsymbol{HFI:CMB:absolute:calibration:100GHz}{$\lsim 2\%$}
\setsymbol{HFI:CMB:absolute:calibration:143GHz}{$\lsim 2\%$}
\setsymbol{HFI:CMB:absolute:calibration:217GHz}{$\lsim 2\%$}
\setsymbol{HFI:CMB:absolute:calibration:353GHz}{$\lsim 2\%$}
\setsymbol{HFI:CMB:absolute:calibration:545GHz}{}
\setsymbol{HFI:CMB:absolute:calibration:857GHz}{}

\setsymbol{HFI:FIRAS:gain:calibration:accuracy:statistical:100GHz}{}
\setsymbol{HFI:FIRAS:gain:calibration:accuracy:statistical:143GHz}{}
\setsymbol{HFI:FIRAS:gain:calibration:accuracy:statistical:217GHz}{}
\setsymbol{HFI:FIRAS:gain:calibration:accuracy:statistical:353GHz}{2.5\%}
\setsymbol{HFI:FIRAS:gain:calibration:accuracy:statistical:545GHz}{1\%}
\setsymbol{HFI:FIRAS:gain:calibration:accuracy:statistical:857GHz}{0.5\%}

\setsymbol{HFI:FIRAS:gain:calibration:accuracy:systematic:100GHz}{}
\setsymbol{HFI:FIRAS:gain:calibration:accuracy:systematic:143GHz}{}
\setsymbol{HFI:FIRAS:gain:calibration:accuracy:systematic:217GHz}{}
\setsymbol{HFI:FIRAS:gain:calibration:accuracy:systematic:353GHz}{}
\setsymbol{HFI:FIRAS:gain:calibration:accuracy:systematic:545GHz}{7\%}
\setsymbol{HFI:FIRAS:gain:calibration:accuracy:systematic:857GHz}{7\%}

\setsymbol{HFI:FIRAS:zero:point:accuracy:100GHz:units}{0.8\MJysr}
\setsymbol{HFI:FIRAS:zero:point:accuracy:143GHz:units}{}
\setsymbol{HFI:FIRAS:zero:point:accuracy:217GHz:units}{}
\setsymbol{HFI:FIRAS:zero:point:accuracy:353GHz:units}{1.4\MJysr}
\setsymbol{HFI:FIRAS:zero:point:accuracy:545GHz:units}{2.2\MJysr}
\setsymbol{HFI:FIRAS:zero:point:accuracy:857GHz:units}{1.7\MJysr}

\setsymbol{HFI:FIRAS:zero:point:accuracy:100GHz}{0.8}
\setsymbol{HFI:FIRAS:zero:point:accuracy:143GHz}{}
\setsymbol{HFI:FIRAS:zero:point:accuracy:217GHz}{}
\setsymbol{HFI:FIRAS:zero:point:accuracy:353GHz}{1.4}
\setsymbol{HFI:FIRAS:zero:point:accuracy:545GHz}{2.2}
\setsymbol{HFI:FIRAS:zero:point:accuracy:857GHz}{1.7}

\setsymbol{HFI:unit:conversion:100GHz:units}{0.2415\MJysrmK}
\setsymbol{HFI:unit:conversion:143GHz:units}{0.3694\MJysrmK}
\setsymbol{HFI:unit:conversion:217GHz:units}{0.4811\MJysrmK}
\setsymbol{HFI:unit:conversion:353GHz:units}{0.2883\MJysrmK}
\setsymbol{HFI:unit:conversion:545GHz:units}{0.05826\MJysrmK}
\setsymbol{HFI:unit:conversion:857GHz:units}{0.002238\MJysrmK}

\setsymbol{HFI:unit:conversion:100GHz}{0.2415}
\setsymbol{HFI:unit:conversion:143GHz}{0.3694}
\setsymbol{HFI:unit:conversion:217GHz}{0.4811}
\setsymbol{HFI:unit:conversion:353GHz}{0.2883}
\setsymbol{HFI:unit:conversion:545GHz}{0.05826}
\setsymbol{HFI:unit:conversion:857GHz}{0.002238}

\setsymbol{HFI:colour:correction:alpha=-2:V1.01:100GHz}{0.9893}
\setsymbol{HFI:colour:correction:alpha=-2:V1.01:143GHz}{0.9759}
\setsymbol{HFI:colour:correction:alpha=-2:V1.01:217GHz}{1.0007}
\setsymbol{HFI:colour:correction:alpha=-2:V1.01:353GHz}{1.0028}
\setsymbol{HFI:colour:correction:alpha=-2:V1.01:545GHz}{1.0019}
\setsymbol{HFI:colour:correction:alpha=-2:V1.01:857GHz}{0.9889}

\setsymbol{HFI:colour:correction:alpha=0:V1.01:100GHz}{1.0008}
\setsymbol{HFI:colour:correction:alpha=0:V1.01:143GHz}{1.0148}
\setsymbol{HFI:colour:correction:alpha=0:V1.01:217GHz}{0.9909}
\setsymbol{HFI:colour:correction:alpha=0:V1.01:353GHz}{0.9888}
\setsymbol{HFI:colour:correction:alpha=0:V1.01:545GHz}{0.9878}
\setsymbol{HFI:colour:correction:alpha=0:V1.01:857GHz}{1.0014}

\begin{document}

%
%\title{WG7\_\projectnumber\ Proposal for an Intermediate Paper}
\title{\papertitle}
%\subtitle{\papertitle}
%This author list corresponds to \title{Author list for PIP 80, Proj. Ref. 7.10: Planck intermediate results. XII. Component separation in the Gould Belt System}
%Prepared by R. Leonardi (rleonardi@sciops.esa.int), ESAC/ESA
%This version is from Thu Mar 07 18:08:28 2013 CET
%\subtitle{There are 181 co-authors in this list}
\author{\small
Planck Collaboration:
P.~A.~R.~Ade\inst{76}
\and
N.~Aghanim\inst{51}
\and
M.~I.~R.~Alves\inst{51}
\and
M.~Arnaud\inst{65}
\and
M.~Ashdown\inst{61, 6}
\and
F.~Atrio-Barandela\inst{17}
\and
J.~Aumont\inst{51}
\and
C.~Baccigalupi\inst{75}
\and
A.~Balbi\inst{31}
\and
A.~J.~Banday\inst{81, 9}
% ...
}
\institute{\small
APC, AstroParticule et Cosmologie, Universit\'{e} Paris Diderot, CNRS/IN2P3, CEA/lrfu, Observatoire de Paris, Sorbonne Paris Cit\'{e}, 10, rue Alice Domon et L\'{e}onie Duquet, 75205 Paris Cedex 13, France\\
\and
Aalto University Mets\"{a}hovi Radio Observatory, Mets\"{a}hovintie 114, FIN-02540 Kylm\"{a}l\"{a}, Finland\\
\and
African Institute for Mathematical Sciences, 6-8 Melrose Road, Muizenberg, Cape Town, South Africa\\
\and
Agenzia Spaziale Italiana Science Data Center, c/o ESRIN, via Galileo Galilei, Frascati, Italy\\
\and
Agenzia Spaziale Italiana, Viale Liegi 26, Roma, Italy\\
\and
Astrophysics Group, Cavendish Laboratory, University of Cambridge, J J Thomson Avenue, Cambridge CB3 0HE, U.K.\\
\and
CITA, University of Toronto, 60 St. George St., Toronto, ON M5S 3H8, Canada\\
\and
CNR - ISTI, Area della Ricerca, via G. Moruzzi 1, Pisa, Italy\\
\and
CNRS, IRAP, 9 Av. colonel Roche, BP 44346, F-31028 Toulouse cedex 4, France\\
\and
California Institute of Technology, Pasadena, California, U.S.A.\\
\and
Centro de Estudios de F\'{i}sica del Cosmos de Arag\'{o}n (CEFCA), Plaza San Juan, 1, planta 2, E-44001, Teruel, Spain\\
\and
Computational Cosmology Center, Lawrence Berkeley National Laboratory, Berkeley, California, U.S.A.\\
\and
Consejo Superior de Investigaciones Cient\'{\i}ficas (CSIC), Madrid, Spain\\
\and
DSM/Irfu/SPP, CEA-Saclay, F-91191 Gif-sur-Yvette Cedex, France\\
\and
DTU Space, National Space Institute, Technical University of Denmark, Elektrovej 327, DK-2800 Kgs. Lyngby, Denmark\\
\and
D\'{e}partement de Physique Th\'{e}orique, Universit\'{e} de Gen\`{e}ve, 24, Quai E. Ansermet,1211 Gen\`{e}ve 4, Switzerland\\
\and
Departamento de F\'{\i}sica Fundamental, Facultad de Ciencias, Universidad de Salamanca, 37008 Salamanca, Spain\\
\and
Departamento de F\'{\i}sica, Universidad de Oviedo, Avda. Calvo Sotelo s/n, Oviedo, Spain\\
\and
Department of Astrophysics/IMAPP, Radboud University Nijmegen, P.O. Box 9010, 6500 GL Nijmegen, The Netherlands\\
\and
Department of Electrical Engineering and Computer Sciences, University of California, Berkeley, California, U.S.A.\\
\and
Department of Physics \& Astronomy, University of British Columbia, 6224 Agricultural Road, Vancouver, British Columbia, Canada\\
\and
Department of Physics and Astronomy, Dana and David Dornsife College of Letter, Arts and Sciences, University of Southern California, Los Angeles, CA 90089, U.S.A.\\
\and
Department of Physics, Gustaf H\"{a}llstr\"{o}min katu 2a, University of Helsinki, Helsinki, Finland\\
\and
Department of Physics, University of California, Santa Barbara, California, U.S.A.\\
\and
Department of Physics, University of Illinois at Urbana-Champaign, 1110 West Green Street, Urbana, Illinois, U.S.A.\\
\and
Dipartimento di Fisica e Astronomia G. Galilei, Universit\`{a} degli Studi di Padova, via Marzolo 8, 35131 Padova, Italy\\
\and
Dipartimento di Fisica e Scienze della Terra, Universit\`{a} di Ferrara, Via Saragat 1, 44122 Ferrara, Italy\\
\and
Dipartimento di Fisica, Universit\`{a} La Sapienza, P. le A. Moro 2, Roma, Italy\\
\and
Dipartimento di Fisica, Universit\`{a} degli Studi di Milano, Via Celoria, 16, Milano, Italy\\
\and
Dipartimento di Fisica, Universit\`{a} degli Studi di Trieste, via A. Valerio 2, Trieste, Italy\\
\and
Dipartimento di Fisica, Universit\`{a} di Roma Tor Vergata, Via della Ricerca Scientifica, 1, Roma, Italy\\
\and
Dipartimento di Matematica, Universit\`{a} di Roma Tor Vergata, Via della Ricerca Scientifica, 1, Roma, Italy\\
\and
Discovery Center, Niels Bohr Institute, Blegdamsvej 17, Copenhagen, Denmark\\
\and
Dpto. Astrof\'{i}sica, Universidad de La Laguna (ULL), E-38206 La Laguna, Tenerife, Spain\\
\and
European Space Agency, ESAC, Planck Science Office, Camino bajo del Castillo, s/n, Urbanizaci\'{o}n Villafranca del Castillo, Villanueva de la Ca\~{n}ada, Madrid, Spain\\
\and
European Space Agency, ESTEC, Keplerlaan 1, 2201 AZ Noordwijk, The Netherlands\\
\and
Helsinki Institute of Physics, Gustaf H\"{a}llstr\"{o}min katu 2, University of Helsinki, Helsinki, Finland\\
\and
INAF - Osservatorio Astronomico di Padova, Vicolo dell'Osservatorio 5, Padova, Italy\\
\and
INAF - Osservatorio Astronomico di Roma, via di Frascati 33, Monte Porzio Catone, Italy\\
\and
INAF - Osservatorio Astronomico di Trieste, Via G.B. Tiepolo 11, Trieste, Italy\\
\and
INAF Istituto di Radioastronomia, Via P. Gobetti 101, 40129 Bologna, Italy\\
\and
INAF/IASF Bologna, Via Gobetti 101, Bologna, Italy\\
\and
INAF/IASF Milano, Via E. Bassini 15, Milano, Italy\\
\and
INFN, Sezione di Bologna, Via Irnerio 46, I-40126, Bologna, Italy\\
\and
INFN, Sezione di Roma 1, Universit`{a} di Roma Sapienza, Piazzale Aldo Moro 2, 00185, Roma, Italy\\
\and
IUCAA, Post Bag 4, Ganeshkhind, Pune University Campus, Pune 411 007, India\\
\and
Imperial College London, Astrophysics group, Blackett Laboratory, Prince Consort Road, London, SW7 2AZ, U.K.\\
\and
Infrared Processing and Analysis Center, California Institute of Technology, Pasadena, CA 91125, U.S.A.\\
\and
Institut N\'{e}el, CNRS, Universit\'{e} Joseph Fourier Grenoble I, 25 rue des Martyrs, Grenoble, France\\
\and
Institut Universitaire de France, 103, bd Saint-Michel, 75005, Paris, France\\
\and
Institut d'Astrophysique Spatiale, CNRS (UMR8617) Universit\'{e} Paris-Sud 11, B\^{a}timent 121, Orsay, France\\
\and
Institut d'Astrophysique de Paris, CNRS (UMR7095), 98 bis Boulevard Arago, F-75014, Paris, France\\
\and
Institute for Space Sciences, Bucharest-Magurale, Romania\\
\and
Institute of Astronomy and Astrophysics, Academia Sinica, Taipei, Taiwan\\
\and
Institute of Astronomy, University of Cambridge, Madingley Road, Cambridge CB3 0HA, U.K.\\
\and
Institute of Theoretical Astrophysics, University of Oslo, Blindern, Oslo, Norway\\
\and
Instituto de Astrof\'{\i}sica de Canarias, C/V\'{\i}a L\'{a}ctea s/n, La Laguna, Tenerife, Spain\\
\and
Instituto de F\'{\i}sica de Cantabria (CSIC-Universidad de Cantabria), Avda. de los Castros s/n, Santander, Spain\\
\and
Jet Propulsion Laboratory, California Institute of Technology, 4800 Oak Grove Drive, Pasadena, California, U.S.A.\\
\and
Jodrell Bank Centre for Astrophysics, Alan Turing Building, School of Physics and Astronomy, The University of Manchester, Oxford Road, Manchester, M13 9PL, U.K.\\
\and
Kavli Institute for Cosmology Cambridge, Madingley Road, Cambridge, CB3 0HA, U.K.\\
\and
Kavli Institute for Theoretical Physics, University of California, Santa Barbara Kohn Hall, Santa Barbara, CA 93106, U.S.A.\\
\and
LAL, Universit\'{e} Paris-Sud, CNRS/IN2P3, Orsay, France\\
\and
LERMA, CNRS, Observatoire de Paris, 61 Avenue de l'Observatoire, Paris, France\\
\and
Laboratoire AIM, IRFU/Service d'Astrophysique - CEA/DSM - CNRS - Universit\'{e} Paris Diderot, B\^{a}t. 709, CEA-Saclay, F-91191 Gif-sur-Yvette Cedex, France\\
\and
Laboratoire Traitement et Communication de l'Information, CNRS (UMR 5141) and T\'{e}l\'{e}com ParisTech, 46 rue Barrault F-75634 Paris Cedex 13, France\\
\and
Laboratoire de Physique Subatomique et de Cosmologie, Universit\'{e} Joseph Fourier Grenoble I, CNRS/IN2P3, Institut National Polytechnique de Grenoble, 53 rue des Martyrs, 38026 Grenoble cedex, France\\
\and
Laboratoire de Physique Th\'{e}orique, Universit\'{e} Paris-Sud 11 \& CNRS, B\^{a}timent 210, 91405 Orsay, France\\
\and
Lawrence Berkeley National Laboratory, Berkeley, California, U.S.A.\\
\and
Max-Planck-Institut f\"{u}r Astrophysik, Karl-Schwarzschild-Str. 1, 85741 Garching, Germany\\
\and
MilliLab, VTT Technical Research Centre of Finland, Tietotie 3, Espoo, Finland\\
\and
National University of Ireland, Department of Experimental Physics, Maynooth, Co. Kildare, Ireland\\
\and
Niels Bohr Institute, Blegdamsvej 17, Copenhagen, Denmark\\
\and
Optical Science Laboratory, University College London, Gower Street, London, U.K.\\
\and
SISSA, Astrophysics Sector, via Bonomea 265, 34136, Trieste, Italy\\
\and
School of Physics and Astronomy, Cardiff University, Queens Buildings, The Parade, Cardiff, CF24 3AA, U.K.\\
\and
Space Sciences Laboratory, University of California, Berkeley, California, U.S.A.\\
\and
Special Astrophysical Observatory, Russian Academy of Sciences, Nizhnij Arkhyz, Zelenchukskiy region, Karachai-Cherkessian Republic, 369167, Russia\\
\and
Stanford University, Dept of Physics, Varian Physics Bldg, 382 Via Pueblo Mall, Stanford, California, U.S.A.\\
\and
UPMC Univ Paris 06, UMR7095, 98 bis Boulevard Arago, F-75014, Paris, France\\
\and
Universit\'{e} de Toulouse, UPS-OMP, IRAP, F-31028 Toulouse cedex 4, France\\
% ...
}

\abstract{We perform an analysis of the diffuse low-frequency Galactic components in the Southern part of the Gould Belt system ($130^\circ\leq l\leq 230^\circ$ and $-50^\circ\leq b\leq -10^\circ$). Strong ultra-violet (UV) flux coming from the Gould Belt super-association is responsible for bright diffuse foregrounds that we observe from our position inside the system and that can help us improve our knowledge of the Galactic emission. Free-free emission and anomalous microwave emission (AME) are the dominant components at low frequencies ($\nu < 40\,$GHz), while synchrotron emission is very smooth and faint. We separate diffuse free-free emission and AME from synchrotron emission and thermal dust emission by using \Planck~data, complemented by ancillary data, using the ``Correlated Component Analysis'' (CCA) component separation method and we compare with the results of cross-correlation of foreground templates with the frequency maps.
We estimate the electron temperature $T_{\rm e}$ from H$\alpha$ and free-free emission using two methods (temperature-temperature plot and cross-correlation) and we obtain $T_{\rm e}$ ranging from  3100 to 5200\,K, for an effective fraction of absorbing dust along the line of sight of 30\% ($f_{\rm d}=0.3$). We estimate the frequency spectrum of the diffuse AME and we recover a peak frequency (in flux density units) of $25.5\pm 1.5$\,GHz. We verify the reliability of this result with realistic simulations that include the presence of biases in the spectral model for the AME and in the free-free template. By combining physical models for vibrational and rotational dust emission and adding the constraints from the thermal dust spectrum from \Planck~and {\it IRAS} we are able to get a good description of the frequency spectrum of the AME for plausible values of the local density and radiation field.}

\keywords{Galaxy: general -- radio continuum: ISM -- radiation mechanisms: general}
\authorrunning{Planck Collaboration et al.}
%\titlerunning{Component separation in the Gould Belt system}
\titlerunning{Galactic diffuse components in the Gould Belt system}
\maketitle
%
%________________________________________________________________

\section{Introduction}
The wide frequency coverage of the \Planck\footnote{\Planck~\ (http://www.esa.int/\Planck~) is a project of the European Space Agency (ESA) with instruments provided by two scientific consortia funded by ESA member states (in particular the lead countries France and Italy), with contributions from NASA (USA) and telescope reflectors provided by a collaboration between ESA and a scientific consortium led and funded by Denmark.} data gives a unique opportunity to study the main four Galactic foregrounds, namely free-free emission, synchrotron emission, anomalous microwave emission (AME) and thermal (vibrational) dust emission.  The different frequency spectra of the components and their different spatial morphologies provide a means for separating the emission components.  In this paper we apply the Correlated Component Analysis method (CCA, \citealt{bonaldi2006}, \citealt{ricciardi2010}), which uses the spatial morphology of the components to perform the separation. 
The local Gould Belt system of current star formation is chosen as a particularly interesting area in which to make an accurate separation of the four foregrounds because of the different morphologies of the components. \cite{gould1879} first noted this concentration of prominent OB associations inclined at 20$^\circ$ to the Galactic plane. It was next identified as an \ion{H}{i} feature (\citealt{davies1960}, \citealt{lindblad1967}).  Along with velocity data from \ion{H}{i} and CO combined with stellar distances from {\it Hipparchos} the total system appears to be a slowly expanding and rotating ring of gas and dust surrounding a system of OB stars within 500\,pc of the Sun \citep{lindblad1997}. A recent modelling of the Gould Belt system by \cite{perrot} gives semi-axes of 373 $\times$ 233\,pc inclined at 17$^\circ$ with an ascending node at $l = 296^\circ$ and a centre 104\,pc distant from us lying at $l = 180^\circ$. The Gould Belt thickness is ~60\,pc. The stars defining the system have ages less than ~$30 \times 10^6$\,yr. 

The free-free emission from ionized hydrogen is well-understood \citep{cliveff}. H$\alpha$ is a good indicator of the emission measure in regions of low dust absorption. Elsewhere a correction has to be applied, which depends on where the absorbing dust lies relative to the H$\alpha$ emission.  The conversion of an emission measure value to a radio brightness temperature at a given frequency requires a knowledge of the electron temperature.  Alternatively, an electron temperature can be derived by assuming a value for the dust absorption. Values for the electron temperature of 4000--8000\,K are found in similar studies (\citealt{banday2003}, \citealt{davies2006}, \citealt{ghosh2011}). Radio recombination line observations on the Galactic plane \citep{alves2011} give values that agree with those of individual \ion{H}{ii} regions, having temperatures that rise with increasing distance from the Galactic centre; the value at the solar distance where the current study applies is 7000--8000\,K. 

The spectrum of synchrotron emission reflects the spectrum of the cosmic-ray electrons trapped in the Galactic magnetic field. At frequencies below a few GHz the brightness temperature spectral index, $\beta_{\rm s}$, is ranging from $-2.5$ to $-2.7$ \citep{broad1989}. Between ~1.0\,GHz and {\it WMAP} and \Planck~frequencies, the spectral index steepens to values from $-2.9$ to $-3.1$ (\citealt{banday2003}, \citealt{davies2006}, \citealt{kogut2011}). 

Thermal dust dominates the Galactic emission at \Planck~frequencies above 100\,GHz. The spectrum is well-defined here with temperature $T_{\rm d}\approx 18$\,K and spectral index $\beta_{\rm d}$ ranging from 1.5 to 1.8 \citep{planck2011-7.0}. In the frequency range 60--143\,GHz the dust emission overlaps that of the free-free emission and AME, making it a critical range for component separation.

The AME component is highly correlated with the far infra-red dust emission (\citealt{kogut1996}, \citealt{leitch1997}, \citealt{banday2003}, \citealt{lagache2003}, \citealt{oliveira2004}, \citealt{fink2004}, \citealt{davies2006}, \citealt{dobler2008}, \citealt{mamd}, \citealt{ysard2010}, \citealt{gold2011}, \citealt{planck2011-7.2}) and is believed to be the result of electric dipole radiation from small spinning dust grains (\citealt{erickson1957}, \citealt{DL1998}) in a range of environments (\citealt{spdust1}, \citealt{ysard-ves2010}). AME is seen in individual dust clouds associated with molecular clouds, photo-dissociation regions, reflection nebulae and \ion{H}{ii} regions  (e.g., \citealt{2002ApJ...566..898F,2004ApJ...617..350F}, \citealt{watson2005}, \citealt{2006ApJ...639..951C,2008MNRAS.391.1075C}, \citealt{dickinson2006,dickinson2007,dickinson2009}, \citealt{scaife2007,scaife2010}, \citealt{2009MNRAS.400.1394A}, \citealt{todorovic2010}, \citealt{murphy2010}, \citealt{planck2011-7.2}, \citealt{dickinson2013}). In the present study we will be examining the AME spectrum in more extended regions.

\begin{figure}
%Fig1
\begin{center}
\includegraphics[width=5.5cm,angle=90]{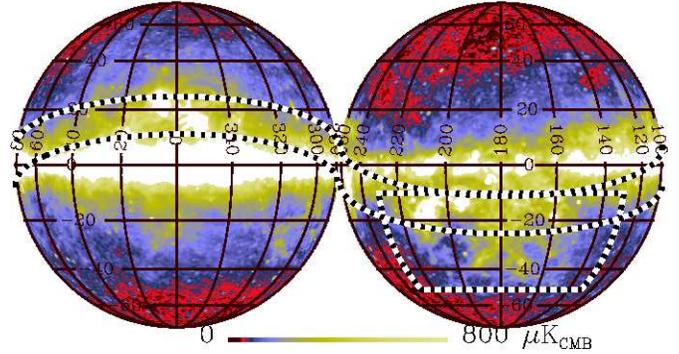}
\caption{Orthographic projection (looking towards the Galactic centre in the left panel and the Galactic anti-centre in the right panel) of the \Planck~CMB-subtracted 30\,GHz channel showing the Gould Belt and the region of interest for this paper (defined by $130^\circ\leq l\leq 230^\circ$ and $-50^\circ\leq b\leq -10^\circ$).}
\label{fig:sam}
\end{center}
\end{figure}

\begin{figure*}
%Fig2
\begin{center}
\includegraphics[width=50mm]{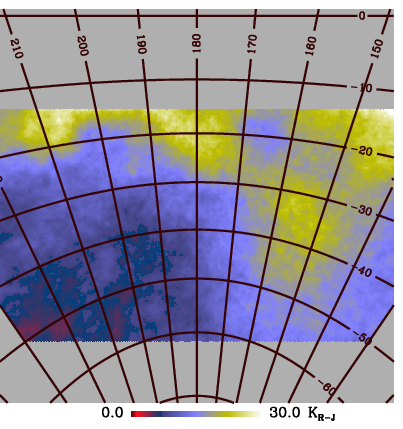}
\includegraphics[width=50mm]{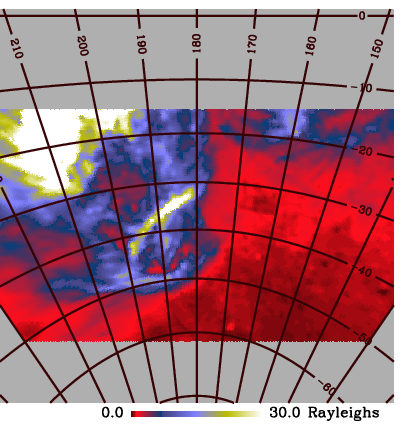}
\includegraphics[width=50mm]{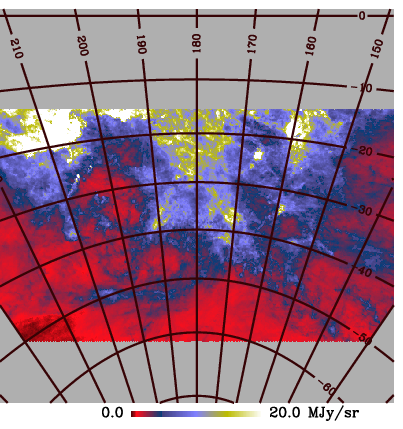}
\includegraphics[width=50mm]{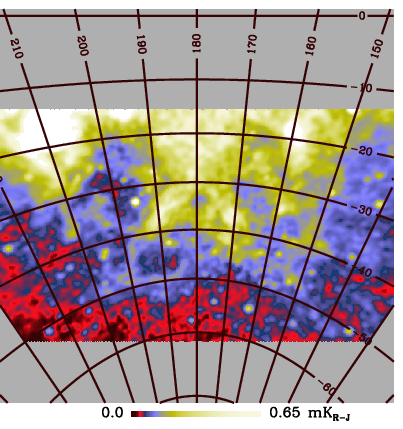}
\includegraphics[width=50mm]{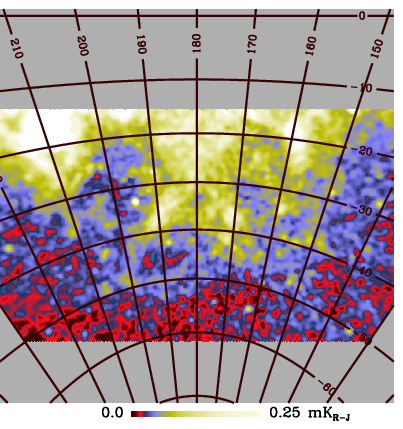}
\includegraphics[width=50mm]{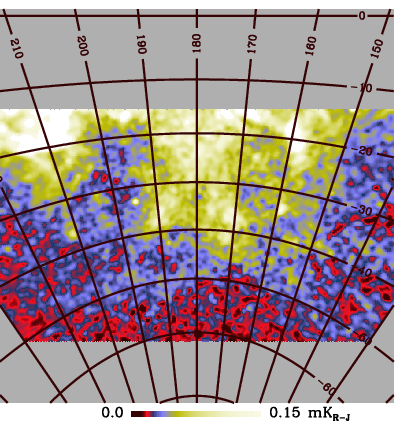}
\includegraphics[width=50mm]{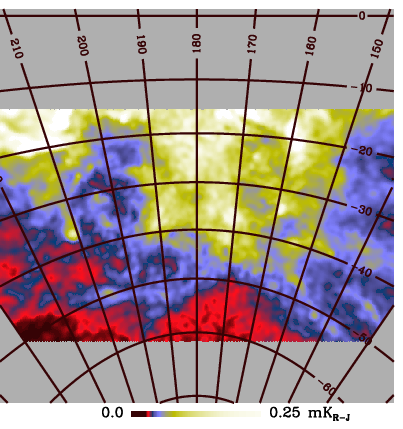}
\includegraphics[width=50mm]{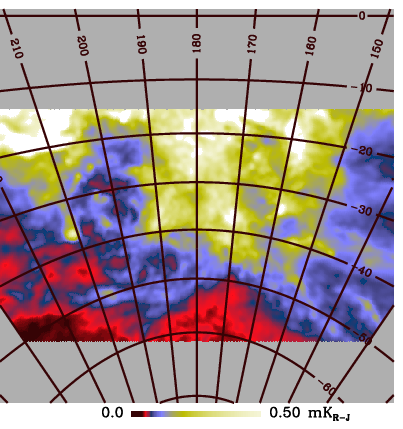}
\includegraphics[width=50mm]{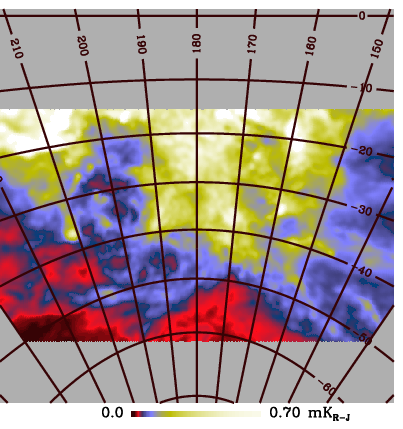}
\caption{Gnomonic projections of the region of interest. {\it Top panels}: \cite{haslam} 408 MHz map ({\it left}); H$\alpha$ map from \cite{cliveff} ({\it middle}); and  100$\microns$ map from \cite{schlegel} ({\it right}) at native resolution. {\it Middle panels} (left to right):  \Planck~CMB-subtracted 30, 44, and 70\,GHz maps at 1$^\circ$ resolution. {\it Bottom panels} (left to right): \Planck~CMB-subtracted 143, 353, and 857\,GHz at 1$^\circ$ resolution.}
\label{fig:GBDR3}
\end{center}
\end{figure*}

\section{Definition of the region of interest and aim of the work}\label{sec:region}

The projection of the Gould Belt disc on the sky is a strip that is superimposed on the Galactic plane, except towards the Galactic centre (Northern Gould Belt) and around $l=180^\circ$ (Southern Gould Belt). In this work we consider the Southern Gould Belt, which can be approximately defined by Galactic coordinates $130^\circ\leq l\leq 230^\circ$ and $-50^\circ\leq b\leq -10^\circ$ (see Fig.~\ref{fig:sam}). This choice gives us a cleaner view of the Gould Belt, because the background emission from the Galactic Plane is weaker here than towards the Galactic centre. Notable structures within the region are the Orion complex, Barnard's arc and the Taurus, Eridanus, and Perseus star-forming complexes. All these emitting regions, including the diffuse emission from the Eridanus shell at $-50^\circ<b<-30^\circ$, are at a distance within 500\,pc from us and thus they belong to the local inter-stellar medium (ISM) associated with the Gould Belt (e.g. \citealt{R&O}, \citealt{boumis2001}). 

In Fig.~\ref{fig:GBDR3} we show the CMB-subtracted \Planck~data at $1^\circ$ resolution, compared with the \cite{haslam} 408 MHz map, which mostly traces the synchrotron component, the \cite{cliveff} H$\alpha$ map, tracing free-free emission, and the 100$\microns$ map from \cite{schlegel}, tracing the dust emission. The visual inspection reveals dust-correlated features at low frequency, which could be attributed to AME. There is also prominent free-free emission, especially strong in the Barnard's arc region (towards $l=207^\circ,b=-18^\circ$). The synchrotron component appears to be sub-dominant with respect to the free-free emission and the AME. 

This work aims to separate and study the diffuse low-frequency foregrounds, in particular AME and free-free emission, in the region of interest. This requires the estimation of the spectral behaviour of the AME (carried out in Sect~\ref{sec:CCA}). We compare this spectrum with predictions for spinning dust emission, one of the mechanisms that is most often invoked to explain AME (Sect.~\ref{sec:ame}).
Having a reconstruction of the free-free emission, we estimate the free-free electron temperature (Sect.~\ref{ff_te}), which relates free-free brightness to emission measure, and investigate the dependence of this result on the dust absorption fraction. 

\section{Description of the analysis}\label{sec:pipeline}

\subsection{Input data} \label{sec:inputs}
\Planck~\citep{tauber2010a, planck2011-1.1} is the third generation
space mission to measure the anisotropy of the cosmic microwave
background (CMB).  It observes the sky in nine frequency bands
covering 30--857\,GHz with high sensitivity and angular resolution
from 31\arcm\ to 5\arcm.  The Low Frequency Instrument (LFI;
\citealt{Mandolesi2010, Bersanelli2010, Planck2011-1.4}) covers the
30, 44, and 70\,GHz bands with amplifiers cooled to 20\,\hbox{K}.  The
High Frequency Instrument (HFI; \citealt{Lamarre2010, Planck2011-1.5})
covers the 100, 143, 217, 353, 545, and 857\,GHz bands with bolometers
cooled to 0.1\,\hbox{K}.  Polarization is measured in all but the
highest two bands \citep{Leahy2010, Rosset2010}.  A combination of
radiative cooling and three mechanical coolers produces the
temperatures needed for the detectors and optics
\citep{planck2011-1.3}.  Two data processing centres (DPCs) check and
calibrate the data and make maps of the sky \citep{planck2011-1.7,planck2011-1.6}.  \Planck's sensitivity, angular resolution, and
frequency coverage make it a powerful instrument for galactic and
extragalactic astrophysics as well as cosmology.  Early astrophysics
results are given in Planck Collaboration VIII--XXVI 2011, based on
data taken between 13~August 2009 and 7~June 2010.  Intermediate
astrophysics results are now being presented in a series of papers
based on data taken between 13~August 2009 and 27~November 2010.

The \Planck~data used throughout this paper are an internal data set known
as DX7, whose properties are described in appendices to the LFI and HFI
data processing papers \citep{NewII, NewVI}.
However, we have tested the analysis to the extent that the results will not change if carried out on the maps which has been released to the public in March 2013.

The specifications of the \Planck~maps are reported in Table~\ref{tab:planckdata}. The dataset used for the analysis consists of full resolution frequency maps and the corresponding noise information. We will indicate whenever the CMB-removed version of this dataset has been used for display purposes. 

When analysing the results we apply a point source mask based on blind detection of sources above  5\,$\sigma$ in each \Planck~map, as described in \cite{planck2011-1.6} and \cite{planck2011-1.7}.
Ancillary data have been used throughout the paper for component separation purposes,  to simulate the sky and data, or to analyse our results. 
The full list of ancillary data is reported in Table~\ref{tab:ancillarydata} with the main specifications.

\begin{table}[tmb]                 % table* is a two-column table.  Drop the * for one column.
\begingroup
\newdimen\tblskip \tblskip=5pt
%\caption{}                          % Caption goes here.
\caption{Summary of \Planck~ data.}
%\label{}                            % Label goes here.
\label{tab:planckdata}
\nointerlineskip
\vskip -3mm
\footnotesize
\setbox\tablebox=\vbox{
   \newdimen\digitwidth 
   \setbox0=\hbox{\rm 0} 
   \digitwidth=\wd0 
   \catcode`*=\active 
   \def*{\kern\digitwidth}
   \newdimen\signwidth 
   \setbox0=\hbox{+} 
   \signwidth=\wd0 
   \catcode`!=\active 
   \def!{\kern\signwidth}
{\tabskip=2em
\halign{ \hfil#\hfil&\hfil#\hfil&\hfil#\hfil\cr                          % Template goes here.
\noalign{\doubleline}
Central frequency    &Instrument&Resolution\cr
[GHz]    &&[arcmin]\cr
\noalign{\vskip 3pt\hrule\vskip 5pt}
*28.5          &\Planck~LFI    &\getsymbol{LFI:FWHM:30GHz:units}  \cr
*44.1          &\Planck~LFI    &\getsymbol{LFI:FWHM:44GHz:units}  \cr
*70.3          &\Planck~LFI   &\getsymbol{LFI:FWHM:70GHz:units}   \cr
100          &\Planck~HFI    &*\getsymbol{HFI:FWHM:Maps:100GHz:units}  \cr
143          &\Planck~HFI    &*\getsymbol{HFI:FWHM:Maps:143GHz:units}   \cr
217          &\Planck~HFI  &*\getsymbol{HFI:FWHM:Maps:217GHz:units}     \cr
353          &\Planck~HFI   &*\getsymbol{HFI:FWHM:Maps:353GHz:units}    \cr
545          &\Planck~HFI   &*\getsymbol{HFI:FWHM:Maps:545GHz:units}     \cr
857          &\Planck~HFI    &*\getsymbol{HFI:FWHM:Maps:857GHz:units}     \cr
\noalign{\vskip 5pt\hrule\vskip 3pt}}}
}
%\endPlancktable                    % ends one-column \halign
\endPlancktablewide                 % ends two-column \halign
%\tablenote a Footnote a.\par
%\tablenote b Footnote b.\par
\endgroup
\end{table}                        % table* is a two-column table.  Drop the * for one column.

\begin{table*}[tmb]                 % table* is a two-column table.  Drop the * for one column.
\begingroup
\newdimen\tblskip \tblskip=5pt
%\caption{}                          % Caption goes here.
\caption{Summary of  ancillary data.}
%\label{}                            % Label goes here.
\label{tab:ancillarydata}
\nointerlineskip
\vskip -3mm
\footnotesize
\setbox\tablebox=\vbox{
   \newdimen\digitwidth 
   \setbox0=\hbox{\rm 0} 
   \digitwidth=\wd0 
   \catcode`*=\active 
   \def*{\kern\digitwidth}
   \newdimen\signwidth 
   \setbox0=\hbox{+} 
   \signwidth=\wd0 
   \catcode`!=\active 
   \def!{\kern\signwidth}
{\tabskip=2em
\halign{\hfil#\hfil&#\hfil&\hfil#\hfil&\hfil#\hfil&\hfil#\cr%#\hfil\cr                          % Template goes here.
%\halign{\hfil#\hfil&#\hfil&\hfil#\hfil&\hfil#\hfil&\hfil#\cr%#\hfil\cr                          % Template goes here.
\noalign{\doubleline}
Central frequency &Label    &Resolution &Reference \cr%  &Usage \cr
%\omit Central frequency &&\omit Label    && \omit Resolution && \omit Reference \cr%  &Usage \cr
[GHz]    &&[arcmin]&&\cr
\noalign{\vskip 3pt\hrule\vskip 5pt}
0.408 &Haslam&60&\cite{haslam}\cr%&1,2,3 \cr
&H$\alpha$&60&\cite{cliveff}\cr%&1,2,3 \cr
&H$\alpha$&6--60&\cite{fink2003}\cr%&3 \cr
22.8--94 &{\it WMAP} 7-yr&56.8--13.8&   \cite{jarosik2010}\cr%&1 \cr
94&&60&\cite{finkbeiner1999}\cr%&2,3\cr 
*2997&100$\microns$&*5&\cite{schlegel}\cr%&2,3\cr
24983, 2997&IRIS Band 1, 4&*4&\cite{IRIS}\cr%&3 \cr
&$E(B-V)$&*5&\cite{schlegel}\cr%&2,3\cr
\noalign{\vskip 5pt\hrule\vskip 3pt}}}
}
%\endPlancktable                    % ends one-column \halign
\endPlancktablewide                 % ends two-column \halign
%\tablenote a Footnote a.\par
%\tablenote b Footnote b.\par
\endgroup
\end{table*}                        % table* is a two-column table.  Drop the * for one column.

\subsection{Components}\label{sec:datamodel}
The main diffuse components present in the data are CMB and Galactic synchrotron emission, free-free emission, thermal dust emission, and anomalous microwave emission (AME). 
The frequency spectrum of the CMB component is well-known: it is accurately described by a black-body having temperature $T_{\rm CMB}=2.7255\,K$ \citep{fixsen2009}.

Thermal dust emission dominates at high frequencies. Its spectral behaviour is  a superposition of modified black-body components identified by temperature $T_{\rm dust}$ and emissivity index $\beta_{\rm d}$: 
\begin{equation}
T_{\rm RJ,dust} (\nu)\propto \nu^{\hbox{\hglue0.7pt}{\beta_{\rm d}+1}}/[\exp(h\nu/kT_{\rm dust})-1] \label{dust},
\end{equation}
where $k$ is the Boltzmann constant and $h$ is the Planck constant.
In the approximation of a single component, over most of the sky we have $T_{\rm dust} \approx 18$\,K and $\beta_{\rm d}$ of 1.5--1.8 (\citealt{finkbeiner1999}, \citealt{planck2011-7.0}, \citealt{planck2011-7.13}).

The frequency spectrum of the free-free component is often described by a power-law with spectral index $-2.14$ in RJ units. A more accurate description \citep[see, e.g.][]{planck2011-7.2} is given by
\begin{equation}
T_{\rm RJ,ff}(\nu)\propto G(\nu) \times (\nu/10)^{-2},\, \label{ff}
\end{equation}
where $G=3.96 (T_4)^{0.21}(\nu/40)^{-0.14}$ is the Gaunt factor, which is responsible for the departure from a pure power-law behaviour. $T_4$ is the electron temperature $T_{\rm e}$ in units of $10^4$\,K ($T_{\rm e}$ can range over 2000--20000\,K, but for most of the ISM it is 4{,}000--15{,}000\,K).

The spectral behaviour of synchrotron radiation can be described to first order by a power-law model with spectral index $\beta_{\rm s}$ that typically assumes values from $-2.5$ to $-3.2$, depending on the position in the sky. Steepening of the synchrotron spectral index with frequency is expected due to energy losses of the electrons. 

The frequency scaling of the AME component is the most poorly constrained. The distinctive feature is a peak around 20--40\,GHz (\citealt{DL1998}, \citealt{dob1}, \citealt{dob2}, \citealt{dob4}). However, a power-law behaviour is compatible with most detections above 23\,GHz (\citealt{banday2003}, \citealt{davies2006}, \citealt{ghosh2011}). This could be the result of a superposition of more peaked components along the line of sight or could indicate a peak frequency lower than 23\,GHz. The most recent {\it WMAP} 9-yr results quote a peak frequency at low latitudes ranging from 10 to 20\,GHz for the spectrum in K$_{\rm R-J}$ units, which means 20--30\,GHz when considering flux density units. 

\subsection{Component separation pipeline}\label{sec:pipeline_compsep}
Several component separation methods adopt the linear mixture data model (see Appendix~\ref{sec:method} for a full derivation). For each line of sight we write: 
\begin{equation}
\vec{x}=\tens{H}\vec{s}+\vec{n}\label{modcca},
\end{equation}
where $\vec{x}$ and $\vec{n}$ contain the data and the noise signals. They are vectors of dimension $N_{\rm d}$, which is the number of frequency channels considered. The vector $\vec{s}$, having dimension $N_{\rm c}$, contains the $N_{\rm c}$ unknown astrophysical components (e.g. CMB, dust emission, synchrotron emission, free-free emission, AME) and the $N_{\rm d} \times N_{\rm c}$ matrix $\tens{H}$, called the mixing matrix, contains the frequency scaling of the components for all the frequencies. The elements of the mixing matrix are computed by integrating the source emission spectra within the instrumental bandpass.
When working in the pixel domain, Eq.~(\ref{modcca}) holds under the assumption that the instrumental beam is the same for all the frequency channels. In the general case, this is achieved by equalizing the resolution of the data maps to the lowest one. When working in the harmonic or Fourier domain, the convolution for the instrumental beam is a multiplication and is linearized without assuming a common resolution.

Within the linear model, we can obtain an estimate $\vec{\hat{s}}$ of the components $\vec{s}$ through a linear mixture of the data: 
\begin{equation}\label{recon}
\vec{ \hat s}=\tens{W}\vec{x},
\end{equation}
where $\tens{W}$ is called the reconstruction matrix. Suitable reconstruction matrices can be obtained from the mixing matrix $\tens{H}$. For example:
\begin{equation}\label{gls}
\tens{W}=[\tens{H}^{T}\tens{C}_{\rm n}^{-1}\tens{H}]^{-1}\tens{H}^{T}\tens{C}_{\rm n}^{-1}
\end{equation}
is called the generalized least square (GLS) solution and only depends on the mixing matrix and on the noise covariance $\tens{C}_{\rm n}$.

The mixing matrix is the key ingredient of component separation. However, as discussed in Sect. \ref{sec:datamodel}, the frequency spectra of the components are not known with sufficient precision to perform an accurate separation. To overcome this problem, our component separation pipeline implements a first step in which the mixing matrix is estimated from the data and a second one in which this result is exploited to reconstruct the amplitudes of the components. 

\subsubsection{Estimation of the mixing matrix}\label{sec:pipeline_cca}
For the mixing matrix estimation we rely on the CCA (\citealt{bonaldi2006}, \citealt{ricciardi2010}), which exploits second-order statistics of the data to estimate the frequency scaling of the components on defined regions of the sky (sky patches). We used the harmonic-domain version of the CCA, whose basic principles of operation are reported in Appendix~\ref{sec:method}. This code works on square sky patches using Fourier transforms. It exploits the data auto- and cross-spectra to estimate a set of parameters describing the frequency scaling of the components. The patch-by-patch estimation prevents the detection of small-scale spatial variations of the spectral properties. On the other hand, by using a large number of samples we retain more information, which provides good constraints, even when the components have similar spectral behaviour. The CCA has been successfully used to separate the synchrotron, free-free and AME components from {\it WMAP} data in \cite{bonaldi2007}. 

We used a patch size of $20^\circ \times 20^\circ$,  obtained as a trade-off between having enough statistics for a robust computation of the data cross-spectra and limited spatial variability of the foreground properties. Given the dimension of the region of interest, we have 10 independent sky patches. However, exploiting a redundant number of patches, widely overlapping with each-other, enables us to eradicate the gaps between them and obtain a result that is independent of any specific selection of patches. We covered the region of interest with patches spaced by $2^\circ$ in both latitude and longitude. By re-projecting the results of the CCA on a sphere and averaging the outputs for each line of sight we can synthesize smooth, spatially varying maps of the spectral parameters (see \citealt{ricciardi2010} for more details). 

\subsubsection{Reconstruction of the component amplitudes}\label{sec:pipeline_gls}

The reconstruction of the amplitudes has been done in pixel space at $1^\circ$ resolution using Eq.~(\ref{gls}), exploiting the output of the previous step. 
To equalize the resolution of the data maps, the $a_{\ell m}$ of each map have been multiplied by a window function, $W^{(\ell)}_S$, given by a $1^\circ$ Gaussian beam divided by the instrumental beam of the corresponding channel (assumed to be Gaussian with full width half maximum (FWHM) as specified in Table 1). This corresponds, in real space, to convolution with a beam $B_{\rm S}$. In order to obtain an estimate of the corresponding noise after smoothing, the noise variance maps should be convolved with $B_{\rm N}=(B_{\rm S})^2$. We did this again in harmonic-space, after having obtained the window function $W^{(\ell)}_{\rm N}$, corresponding to $B_{\rm N}$, by Legendre transforming $W^{(\ell)}_{\rm S}$, squaring the result, and Legendre transforming back. 

The smoothing process also correlates noise between different pixels, which means that the RMS per pixel obtained as detailed above is not a complete description of the noise properties. However, the estimation of the full covariance of noise (and its propagation through the separation in Eqs.~\ref{recon} and \ref{gls}) is
very computationally demanding. In this work we take into account only the diagonal noise covariance and neglect any correlation between noise in different pixels. In a signal-dominated case, such as the one considered here, the errors on the noise model have very small impact on the results.

\section{AME frequency spectrum}\label{sec:CCA}
\begin{figure*}
%fig3
\begin{center}
\begin{tabular}{cc}
\includegraphics[width=8.5cm]{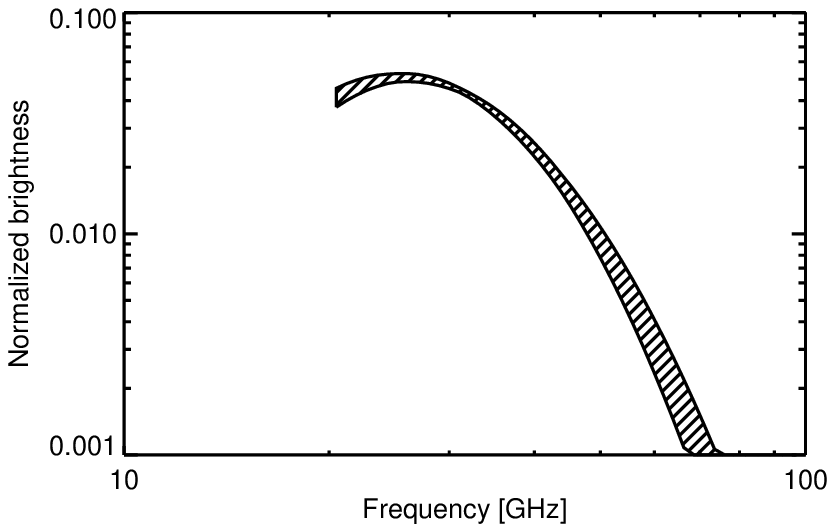}&\includegraphics[width=8.5cm]{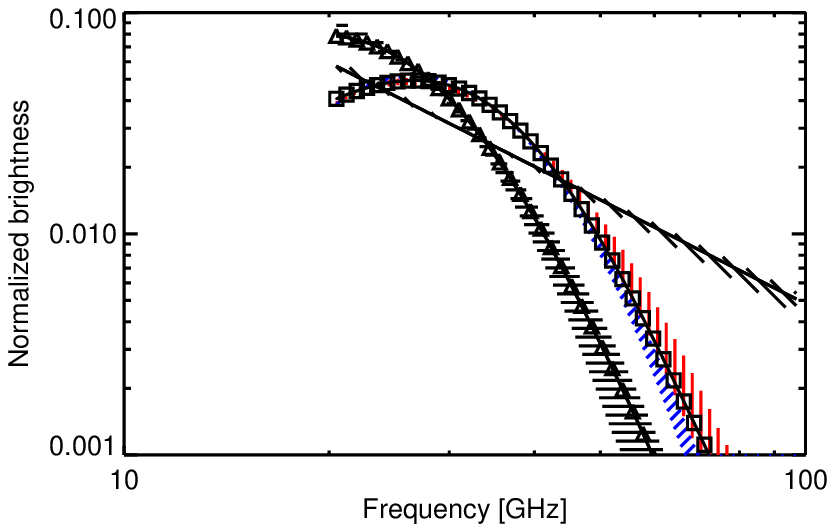}\\
\includegraphics[width=8.5cm]{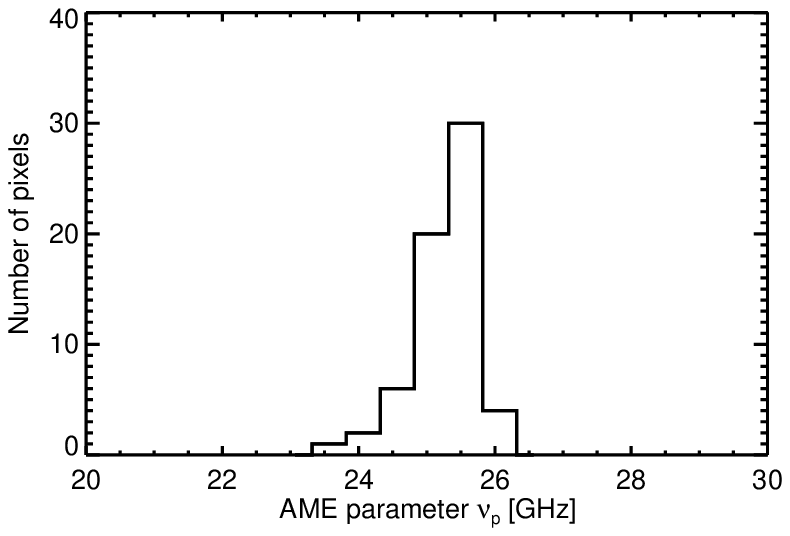}&\includegraphics[width=8.5cm]{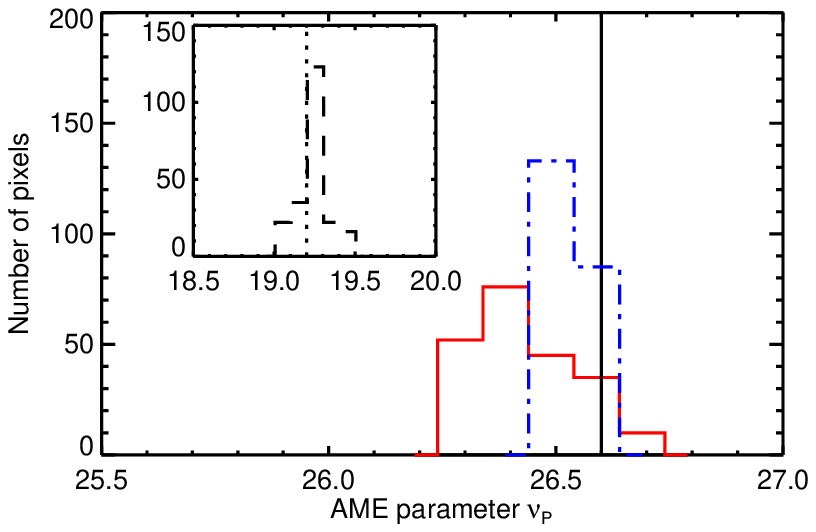}\\
\includegraphics[width=8.5cm]{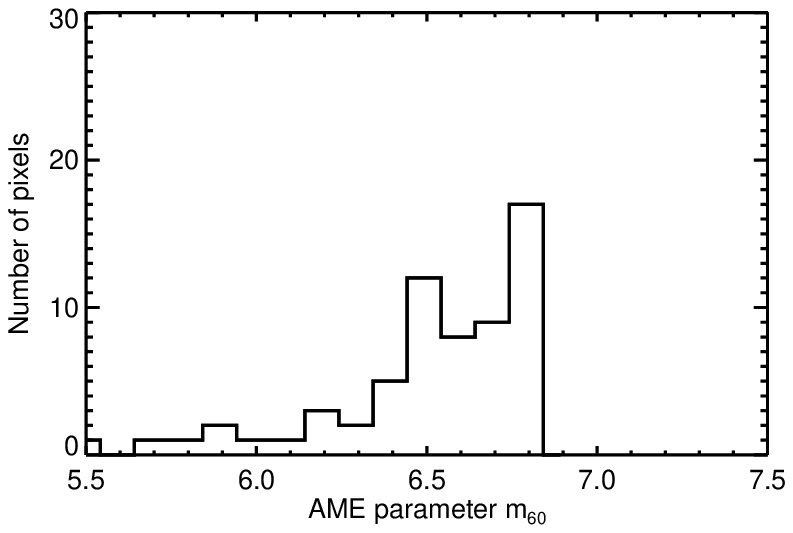}&\includegraphics[width=8.5cm]{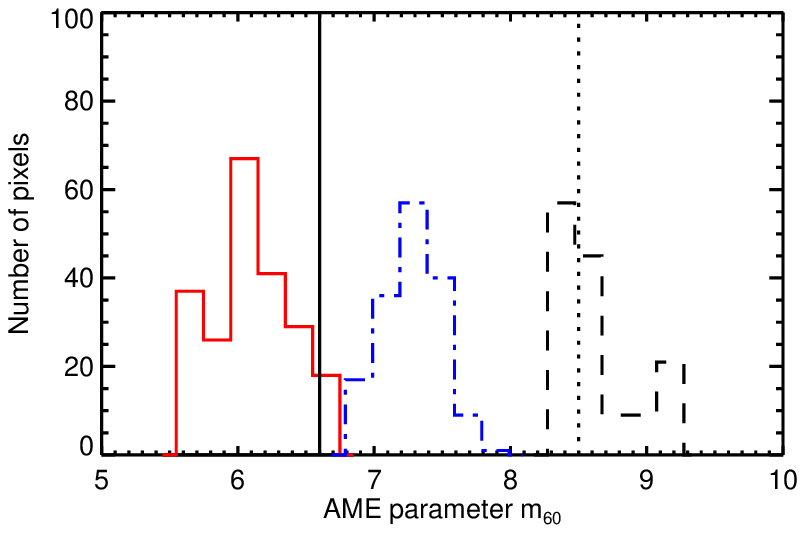}\\
\end{tabular}
\caption{CCA estimation of the AME frequency spectrum in the region of interest for real data ({\it left panels)} and simulated data ({\it right panels}). {\it Top}: estimated spectra including 1$\sigma$ errors. {\it Middle and bottom}: histograms of the spectral parameters $m_{60}$ and $\nu_{\rm p}$ on $N_{\rm side}=16$ estimated spectral index maps. For the simulated case ({\it right panels}) we consider two convex spectra peaking at 19\,GHz and 26\,GHz and a power-law model.  {\it Top right}: the true inputs are shown as solid black lines (power-law) with triangles (19\,GHz peak) and squares (26\,GHz peak) and the estimates as shaded areas. The blue and red colours show estimates done exploiting the free-free templates FF$_{\rm{1}}$ and FF$_{\rm{2}}$ described in Appendix~\ref{sec:simulations}. {\it Middle and bottom right panels}: the true inputs are shown by solid and dotted vertical lines for the simulations peaking at 26\,GHz and 19\,GHz respectively; the blue dot-dashed and red solid histograms show the estimates obtained using the FF$_{\rm{1}}$ and FF$_{\rm{2}}$ templates, and the black dashed lines show the estimates for the 19\,GHz input spectrum.} 
\label{fig:amespec_data}
\end{center}
\end{figure*}
We modelled the mixing matrix to account for five components: CMB; synchrotron emission; thermal dust emission; free-free emission; and AME. We neglected the presence of the CO component by excluding from the analysis the 100 and 217\,GHz \Planck~channels, which are significantly contaminated by the CO lines $J=1\to 0$ and $J=2 \to 1$ respectively \citep{planck2011-1.7}. CO is also present at 353\,GHz, where it can contaminate the dust emission by up to 3\% in the region of interest, and at 545 and 857\,GHz, where the contamination is negligible. 
For the estimation of the mixing matrix we used the following dataset:
\begin{itemize}
\item \Planck~30, 44, 70, 143 and 353\,GHz channels;
\item {\it WMAP} 7-yr K band (23\,GHz); 
\item Haslam et al. 408\,MHz map;
\item Predicted free-free emission at 23\,GHz based on the H$\alpha$ \cite{cliveff} template corrected for dust absorption with the \cite{schlegel} $E(B-V)$ map by assuming a dust absorption fraction of 0.33. 
\end{itemize} 
We verified that the inclusion of the {\it WMAP} Ka--W bands in this analysis did not produce appreciable changes in the results. The explored frequency range is now covered by \Planck~data with higher angular resolution and sensitivity. 
Caution is needed when using H$\alpha$ as a free-free tracer: dust absorption \citep{cliveff} and scattering of H$\alpha$ photons from dust grains (\citealt{woodsrey}, \citealt{dob3}) cause dust-correlated errors in the free-free template, which could bias the AME spectrum. The impact of such biases has been assessed through simulations as described in Sect.~\ref{sec:simul_0}. 

For dust emission we used the model of Eq.~(\ref{dust}) with $T_{\rm d}=18$\,K and estimated the dust spectral index $\beta_{\rm d}$. The reason why we fixed the dust temperature is that this parameter is mostly constrained by high-frequency data, which we do not include in this analysis. 
In fact, a single modified black-body model with constant $\beta_d$ does not provide a good description of the dust spectrum across the frequency range covered by \Planck. In particular, $\beta_d$ results to be flatter in the microwaves ($\nu \leq 353$\,GHz) compared to the millimetre ($\nu > 353$\,GHz).

The temperature $T_{\rm d}=18$\,K we are adopting is consistent with the 1-component dust model by \cite{finkbeiner1999} and in good agreement with the median temperature of 17.7\,K estimated at $|b|>10^\circ$ by \cite{planck2011-7.0}. For the dust spectral index we obtained $\beta_{\rm d}=1.73 \pm 0.09$. 
For synchrotron radiation we adopted a power-law model with fixed spectral index $\beta_{\rm s}=-2.9$ \citep[e.g.,][]{mamd}, as the weakness of the signal prevented a good estimation of this parameter. We verified that different choices for $\beta_{\rm s}$ (up to a 10\,\% variation, $\beta_{\rm s}$ from -2.6 to -3.2) changed the results for the other parameters only of about 1\,\%, due to the weakness of the synchrotron component with respect to AME and thermal dust. 
As a spectral model for AME we adopted the best-fit model of \cite{bonaldi2007}, which is a parabola in the $\log (S)$-$\log (\nu)$ plane parametrized in terms of peak frequency $\nu_{\rm p}$\footnote{The peak frequency $\nu_{\rm p}$ is defined for the specrum in flux density units.} and slope at 60\,GHz $m_{60}$:
\begin{equation}
\log T_{\rm RJ,AME}(\nu)  \propto  \left(\frac{m_{60}\log
\nu_{p}}{\log(\nu_{\rm p}/60)}+2\right)\log \nu + \frac{m_{60}(\log\nu)^2}{2\log(\nu_{\rm p}/{60})} \label{tspin}.
\end{equation}
Details of the model and justification of this choice are given in Appendix~\ref{sec:appa}. 
We also tested a pure power-law model ($T_{\rm RJ,AME}(\nu) \propto \nu^\alpha$) for AME, fitting for the spectral index $\alpha$, but we could not obtain valid estimates in this case. This is what we expect when the true spectrum presents some curvature, as verified through simulations (see Sect.~\ref{sec:simul_0} and Appendix~\ref{sec:simulations}). 

Our results for the AME spectrum are shown in the left panels of Fig.~\ref{fig:amespec_data}.
On average, the AME peaks at 25.5\,GHz, with a standard deviation of 0.6\,GHz, which is within estimation errors (1.5\,GHz). This means we find no significant spatial variations of the spectrum of the AME in the region of the sky considered here. However, we recall that this only applies to diffuse AME, as our pipeline cannot detect small-scale spatial variations, and we are restricted to a limited area of the sky. 

Our results on the peak frequency of the AME are similar to those of \cite{planck2011-7.2} for Perseus and $\rho$ Ophiuchi. {\it WMAP} 9-yr MEM analysis \citep{bennett2012} measures the position of the peak for the spectrum in K$_{\rm R-J}$ units and finds a typical value of 14.4\,GHz for diffuse AME at low latitudes, which roughly corresponds to 27\,GHz when the spectrum is in flux density units. According to previous work, at higher latitudes the peak frequency is probably lower (see e.g. \citealt{banday2003}, \citealt{davies2006}, \citealt{ghosh2011}). Interestingly, the same CCA method used in this paper yields $\nu_p$ around 22\,GHz when applied to the North Celestial Pole region (towards $l=125^\circ$, $b=25^\circ$, \citealt{special}). Spatial variations of the physical properties of the medium could explain these differences. 

In the hypothesis of spinning dust emission, there are many ways to achieve a shift in the peak frequency. As the available data do not allow us to discriminate between them, we will just mention two main possibilities.  
The first is a change in the density of the medium, lower densities being associated with lower peak frequencies (see also Table~\ref{tab:commontab}). Indeed the AME spectrum is modelled with densities of 0.2--0.4\,cm$^{-3}$ in \cite{special}, while it requires higher densities in the Gould Belt region, as discussed in Sect.~\ref{sec:ame}. The second possibility is a change in the size distribution of the dust grains, smaller sizes yielding higher peak frequencies. We will return to these aspects in Sect.~\ref{sec:ame}.

\subsection{Assessment through simulations}\label{sec:simul_0}
The reliability of our results has been tested with simulations.
The main purposes of this assessment are:
\begin{itemize}
\item to verify the ability of our procedure to accurately recover the AME spectrum for different input models; 
\item to investigate how the use of foreground templates --- free-free in particular --- can bias the results. 
\end{itemize}
We did this by applying the procedure described in Sect.~\ref{sec:CCA} to sets of simulated data, for which the true inputs are known. For the first target, we performed three separate simulations including a different AME model: two spinning dust models, peaking at 19\,GHz and 26\,GHz, and a spatially varying power-law. For the second target, we introduced dust-correlated biases in the free-free template and quantified their impact on the estimated parameters. The full description of the simulations and of the tests performed is given in Appendix~\ref{sec:simulations}.

The results are displayed in the right panels of Fig.~\ref{fig:amespec_data}. 
In the top panel we show, for each of the three tested input models, the true spectrum (solid line) and the estimated spectrum with errors (shaded area). 
The red and blue areas distinguish between two free-free templates (referred to as FF$_{\rm{1}}$ and FF$_{\rm{2}}$), which are biased in a different way with respect to the simulated free-free component. In the middle and bottom panels we show the histograms of the recovered spectral parameters compared with the true inputs (vertical lines); the red and blue colours are as before.
We conclude the following:
\begin{itemize}
\item If the input AME is a convex spectrum, we are able to accurately recover the peak frequency, $\nu_{\rm p}$, for both the 19 and 26\,GHz input values. Our pipeline is able to distinguish very clearly between the two input models; biases in the free-free template do not affect the recovery of the peak frequency.
\item The estimated spectrum can be slightly biased above 40--50\,GHz, where the AME is faint, as a result of limitations of the spectral mode we are using (see Appendix~\ref{sec:appa}) and errors in the free-free template. The systematic error on $m_{60}$ is quantified as 0.5--0.6.
\item If the input AME spectrum is a power-law, we obtain a good recovery when fitting for a spectral index.
\end{itemize}

When the AME is a power-law the parabolic model is clearly wrong, as the parameter describing the position of the peak is completely unconstrained and the model steepens considerably with frequency. Similarly, when the AME is a curved spectrum the power-law model is too inaccurate to describe it. As expected, both these estimations fail to converge. 
We note that the distribution of $m_{60}$ recovered on real data is quite different from that obtained from the simulation. This could indicate spatial variability of the true spectrum, which is not included in the simulation. It could also indicate that the systematic errors on $m_{60}$ predicted by simulations, as we just described, are different in different regions of the sky, thus creating a non-uniform effect.  

\section{Reconstruction of the amplitudes}\label{sec:GLS}
%fig4
\begin{figure*}
\begin{center}
\includegraphics[width=4cm]{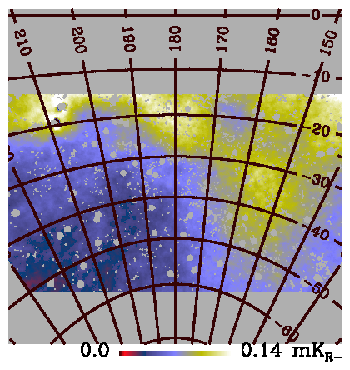}
\includegraphics[width=4cm]{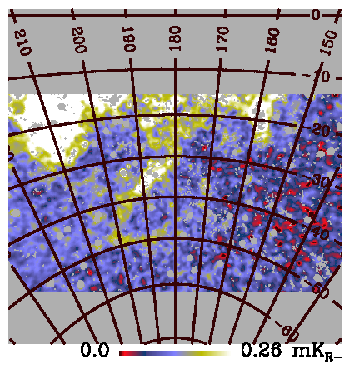}
\includegraphics[width=4cm]{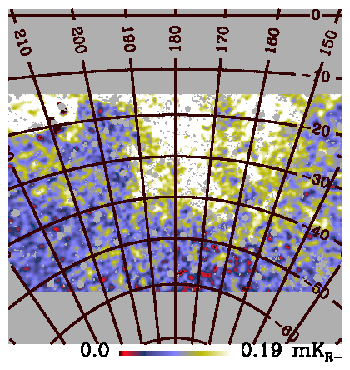}
\includegraphics[width=4cm]{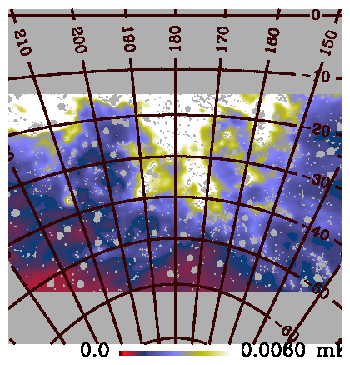}
\includegraphics[width=4cm]{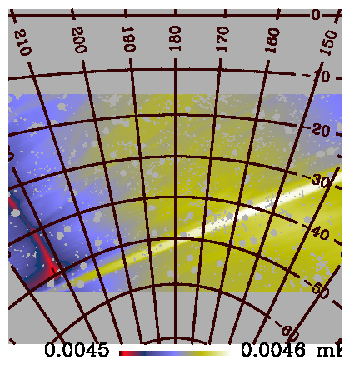}
\includegraphics[width=4cm]{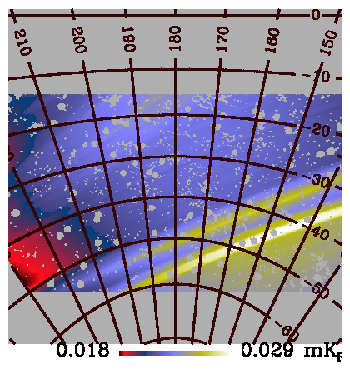}
\includegraphics[width=4cm]{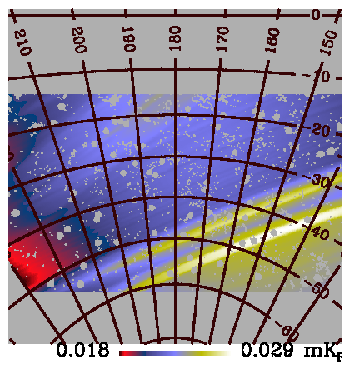}
\includegraphics[width=4cm]{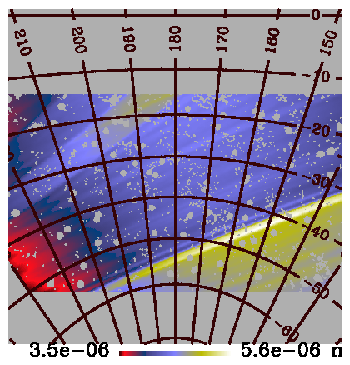}
\includegraphics[width=4cm]{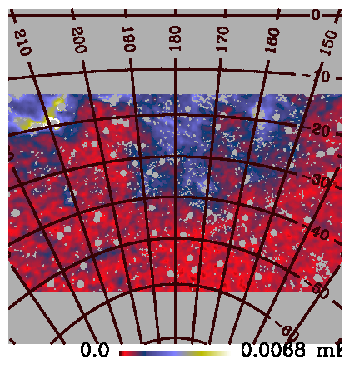}
\includegraphics[width=4cm]{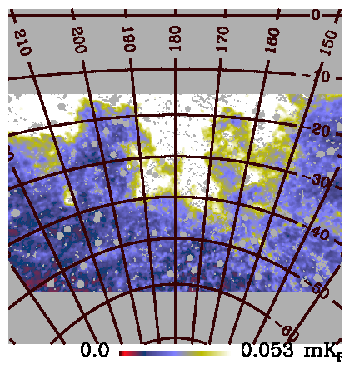}
\includegraphics[width=4cm]{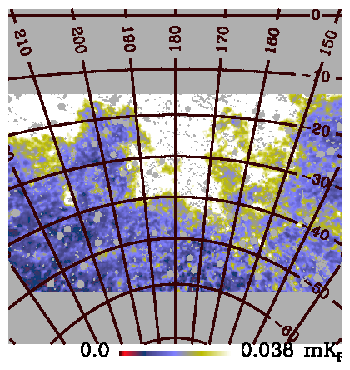}
\includegraphics[width=4cm]{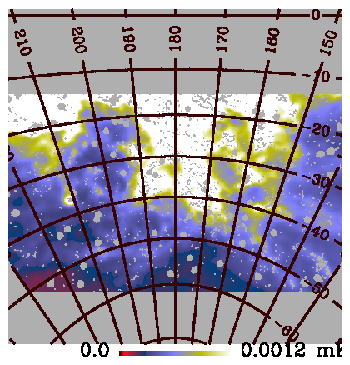}
\includegraphics[width=4cm]{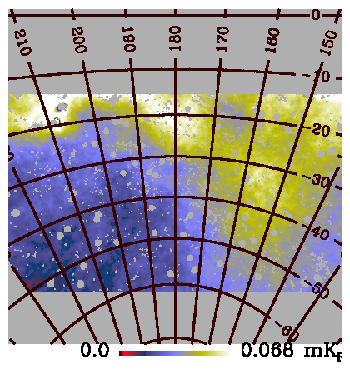}
\includegraphics[width=4cm]{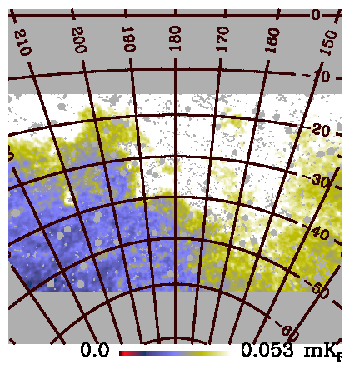}
\includegraphics[width=4cm]{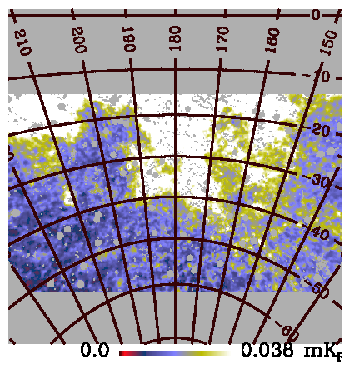}
\includegraphics[width=4cm]{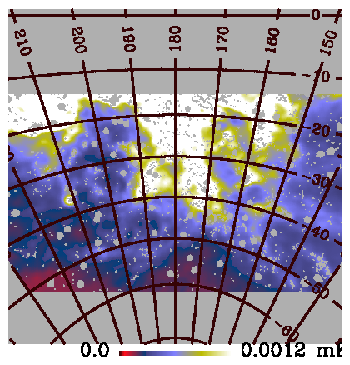}

\caption{1$^\circ$ resolution reconstruction at 30 GHz of ({\it from left to right}): synchrotron emission; free-free emission; AME; and thermal dust emission. These reconstructions are performed as described in Sect.~\ref{sec:pipeline_gls}. {\it Rows from top to bottom}: component amplitudes; noise RMS; predicted RMS of component separation error due to the estimation of AME and thermal dust spectra;  and predicted RMS of component separation error including a random error on $\beta_{\rm s}=-2.9 \pm 0.1$.}
\label{fig:GLS}
\end{center}
\end{figure*}
\begin{figure*}
%fig5
\begin{center}
\begin{tabular}{cc}
\includegraphics[width=8.8cm]{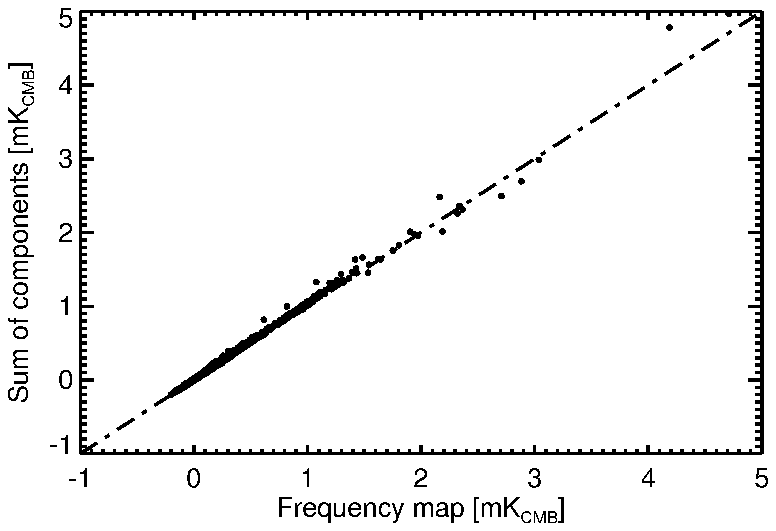}&\includegraphics[width=8cm]{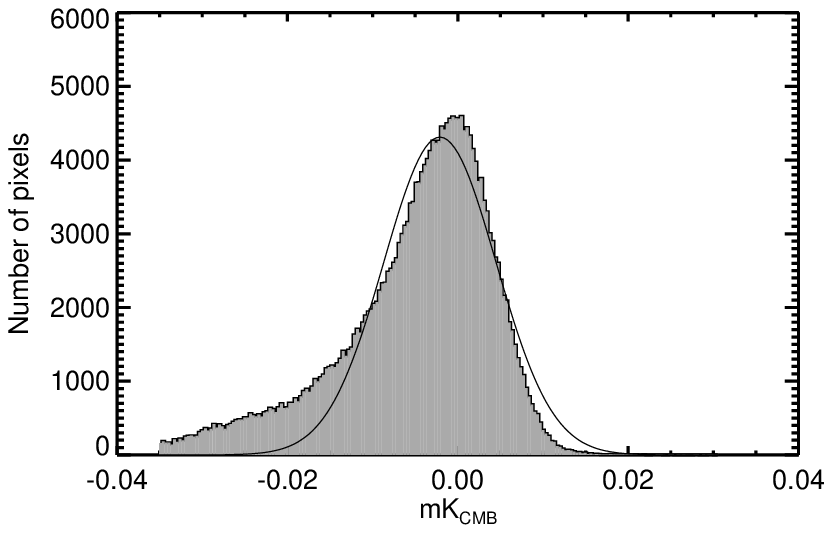}\\
\includegraphics[width=8.8cm]{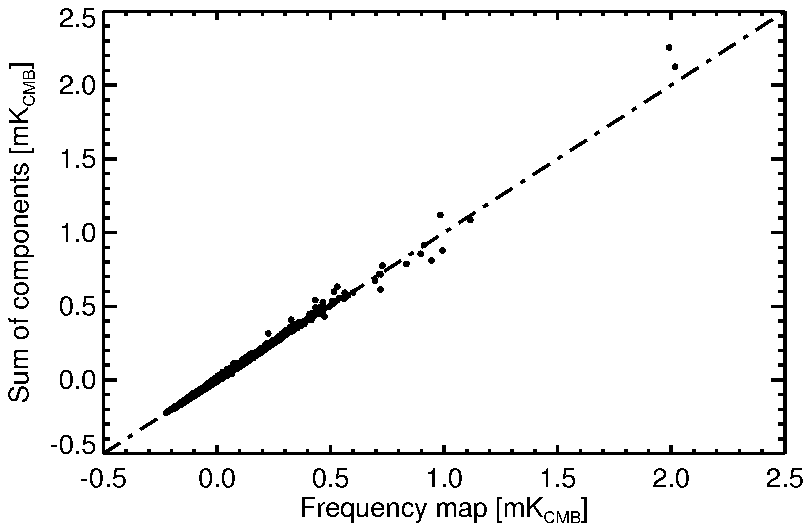}&\includegraphics[width=8cm]{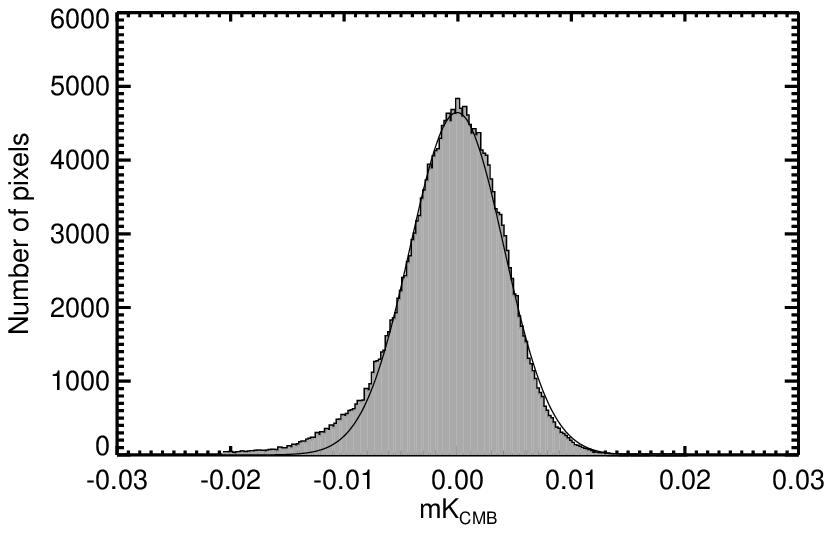}\\
\includegraphics[width=8.8cm]{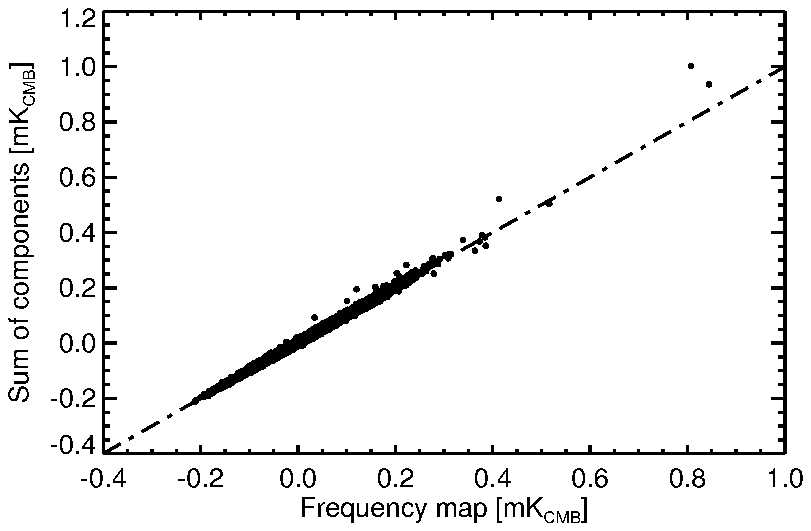}&\includegraphics[width=8cm]{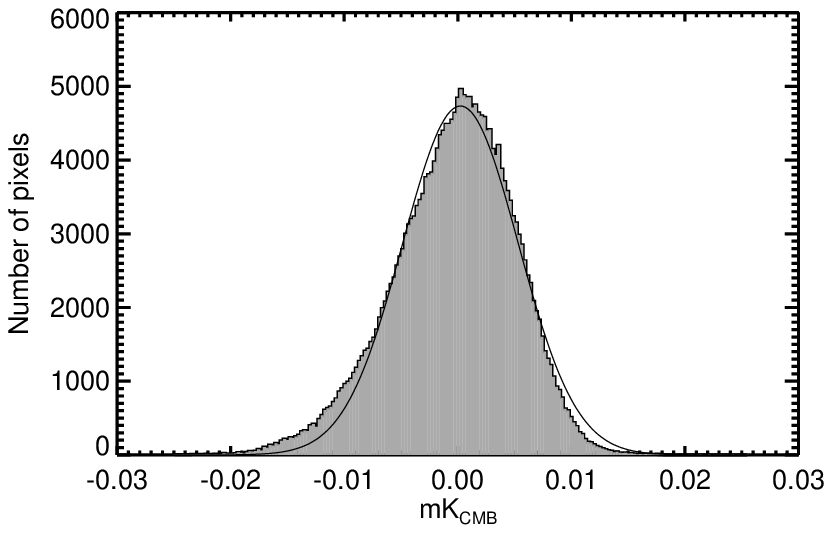}\\
\end{tabular}
\caption{Validation of the reconstructed components shown in Fig.~\ref{fig:GLS}. {\it Left panels}: sum of the components vs frequency maps at ({\it from top to bottom}) 30, 44, and 70\,GHz. The line is the $x=y$ relation. {\it Right panels}: pixel distribution of the residual (frequency map-sum of the components) maps compared to the best-fit Gaussian distribution.}
\label{fig:residuals}
\end{center}
\end{figure*}
The reconstruction of the amplitude of the components has been performed on the $1^\circ$ resolution version of the dataset. We used the same frequencies exploited for the estimation of the mixing matrix, except for the free-free template, which has been excluded to avoid possible biases in the reconstruction. 
The results are shown in  Fig.~\ref{fig:GLS}.
The first and second rows show the components reconstructed at 30\,GHz (from left to right: synchrotron emission, free-free emission, AME, and thermal dust emission) and the corresponding noise RMS maps. Thanks to the linearity of the problem, the noise variance maps can be obtained by combining the noise variance maps of the channels at $1^\circ$ degree resolution with the squared reconstruction matrix $\tens{W}$.
The noise on the synchrotron and thermal dust maps is low compared to that for free-free and AME. This is because the 408\,MHz map and the \Planck~353\,GHz channel give good constraints on the amplitudes of synchrotron and thermal dust emission respectively. 

The AME component is correlated at about 60\,\% and 70\,\% with the 100$\microns$ and the $E(B-V)$ dust templates by \cite{schlegel}, 40\,\% with \cite{haslam} 408 MHz and 20\,\% with H$\alpha$. This favours emission mechanisms based on dust rather than to other hypotheses, such as curved synchrotron emission and free-free emission. The $E(B-V)$ template correlates better with thermal dust emission than the 100$\microns$ map (the correlation coefficients being $0.73\pm 0.01$ and $0.96\pm 0.01$ respectively). This is expected if AME is dust emission. In fact, both spinning dust and thermal dust emission are proportional to the column density, for which $E(B-V)$ is a better estimator than the 100$\microns$ emission, which is strongly affected by the dust temperature.
\begin{figure}
%%fig6
\begin{center}
\includegraphics[width=88mm]{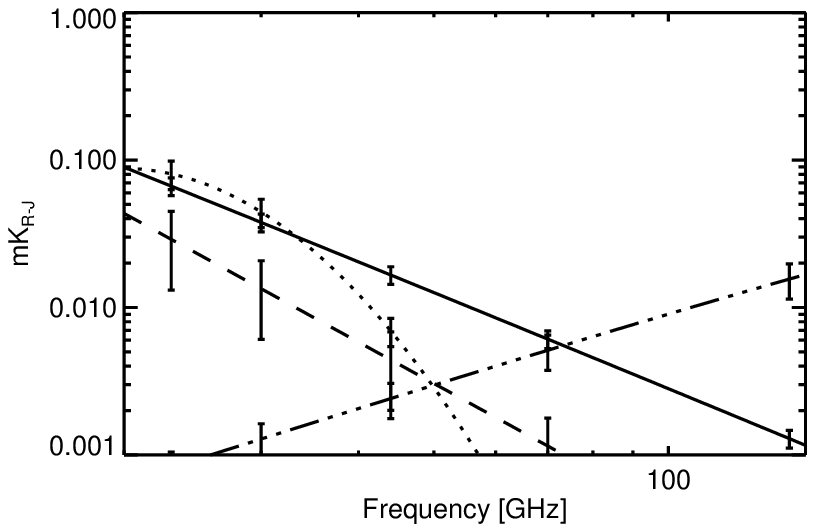}
\includegraphics[width=88mm]{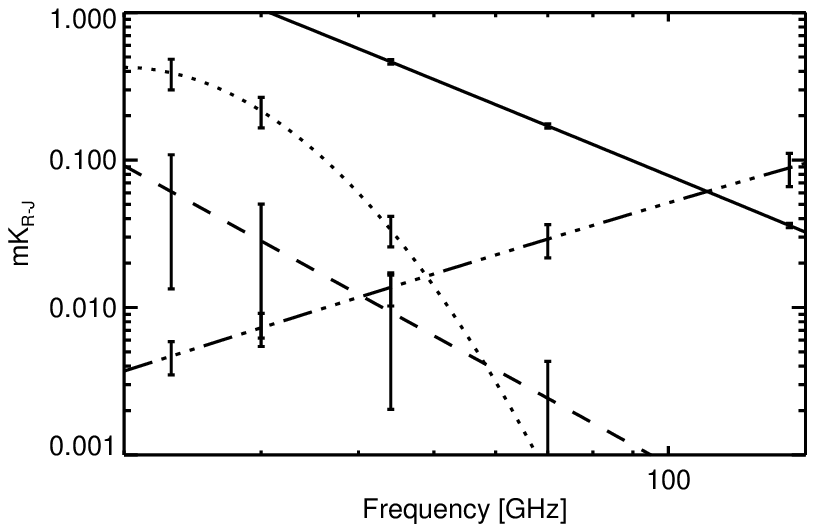}
\caption{Estimated spectra of synchrotron emission (dashed line), free-free emission (solid line), thermal dust emission (dash-dotted line) and AME (dotted line) for average local properties ({\it top}) and for Barnard's region ({\it bottom}). \label{fig:GLS_spec}}
\end{center}
\end{figure}

The errors due to the separation process (third and fourth row of Fig.~\ref{fig:GLS}) are obtained by propagating via Monte-Carlo the uncertainties on the mixing matrix estimated by CCA to the reconstruction of the components (see \citealt{ricciardi2010} for more details). Essentially, the mixing matrix parameters are randomized according to their posterior distributions; the component separation error on the amplitudes is estimated as the variance of GLS reconstructions for different input mixing matrices. 

One complication is that in the present analysis we did not estimate the synchrotron spectral index, but we fixed it at $\beta_{\rm s}=-2.9$. 
Thus, we do not have errors on the synchrotron spectral index from our analysis. We therefore considered two cases: one in which we propagated only the errors on the AME and thermal dust spectral parameters, thus assuming no error on $\beta_{\rm s}$ (third row of Fig.~\ref{fig:GLS}); and another in which we included an indicative random error $\Delta \beta_{\rm s}=0.1$ (last row of Fig.~\ref{fig:GLS}).  

The predicted error due to separation is generally higher than noise and on average of the order 15--20\,\% of the component amplitude for AME, free-free, and dust. Once we allow some scatter on $\beta_{\rm s}$, the predicted error on synchrotron emission becomes of the order of 50\,\%: this indicates that the reconstruction of this component is essentially prior-driven. The inclusion of  $\Delta \beta_{\rm s}$ has some effect on the error prediction for free-free emission, while AME and dust are mostly unaffected.

To evaluate the quality of the separation we compared the frequency maps with the sum of the reconstructed components at the same frequency. 
In the left panels of Fig.~\ref{fig:residuals} we plot the sum of the components for 30, 44, and 70\,GHz against the amplitude of the frequency map. The comparison is made at $1^\circ$ resolution with $N_{\rm{side}}=128$ pixels.
The dashed line indicates the $x=y$ relation, which corresponds to the ideal case in which the two maps are identical. 

The agreement between data and predictions is in general very good.
The scatter of the points does not measure the quality of the separation but the signal-to-noise of the maps. It increases from 30 to 70\,GHz, as the foreground signal gets weaker. The errors in the component separation show up as systematic departures of the data from the prediction. As those are not apparent, we also show on the right panels of Fig.~\ref{fig:residuals} the pixel distribution of the residual map compared to the best-fit Gaussian distribution. At 44 and 70\,GHz the scatter, though quite small, dominates the residual and covers the systematic effects, with the exception of a few outliers, mostly due to compact sources. At 30\,GHz the scatter is low enough to reveal a feature: a sub-sample of pixels in which the reconstructed signal is higher than the true one, thus creating a negative in the residual. 

This kind of systematic effect is very difficult to avoid when separating many bright components, because small errors in the mixing matrix cause bright features in the residual maps. Our Monte-Carlo approach is however able to propagate these errors.  
At 30\,GHz the brightest components are AME and free-free emission, for which the predicted component separation error is on average 0.04--0.05\,mK$_{\rm CMB}$, in agreement with the level of the non-Gaussian residuals. 
Coherent structures in the residual maps are induced by the low resolution of the maps of spectral parameters, which means that over nearby pixels the error in the mixing matrix, and thus on the separation, is similar.

In Fig.~\ref{fig:GLS_spec} we show the amplitude of the components as a function of frequency. The top panel represents the typical behaviour in the Gould Belt, while the bottom one refers to a particular case, Barnard's region where free-free emission is particularly strong. The points are the average amplitude of the components at each frequency within the selected regions of the sky. 
The scaling of the amplitudes with frequency is, by construction, given by the spectral model estimated with CCA. The error bars measure the scatter induced on the amplitudes by the errors on the spectral parameters (also including $\Delta \beta_{\rm s}=0.1$). 

\section{Free-free electron temperature}\label{ff_te}
The intensity of the free-free emission at a given frequency with respect to H$\alpha$ can be expressed as
\begin{equation}
\frac{T_{\rm ff}(\nu)[\mu {\rm K}_{\rm R-J}]}{\rm{H}{\alpha}[\rm Rayleighs]}=14\, T_4^{0.517} \times 10^{0.029/T_4}\times 1.08 \, G(\nu)(\nu/10)^{-2},
\label{ff2halpha}
\end{equation}
where $\rm G(\nu)$ is the Gaunt factor already introduced in Sect.~\ref{sec:datamodel} and $T_4$  is the electron temperature $T_{\rm e}$ in units of $10^4$\,K. 
In the previous equation, H$\alpha$ has been corrected for dust absorption. Following \citep{cliveff}, the correction depends on $f_{\rm d}$, the effective dust fraction in the line of sight actually absorbing the H$\alpha$. Therefore $f_{\rm d}$ and $T_{\rm e}$ are degenerate parameters. 

\begin{figure}
%fig7
\begin{center}
\includegraphics[width=6.5cm]{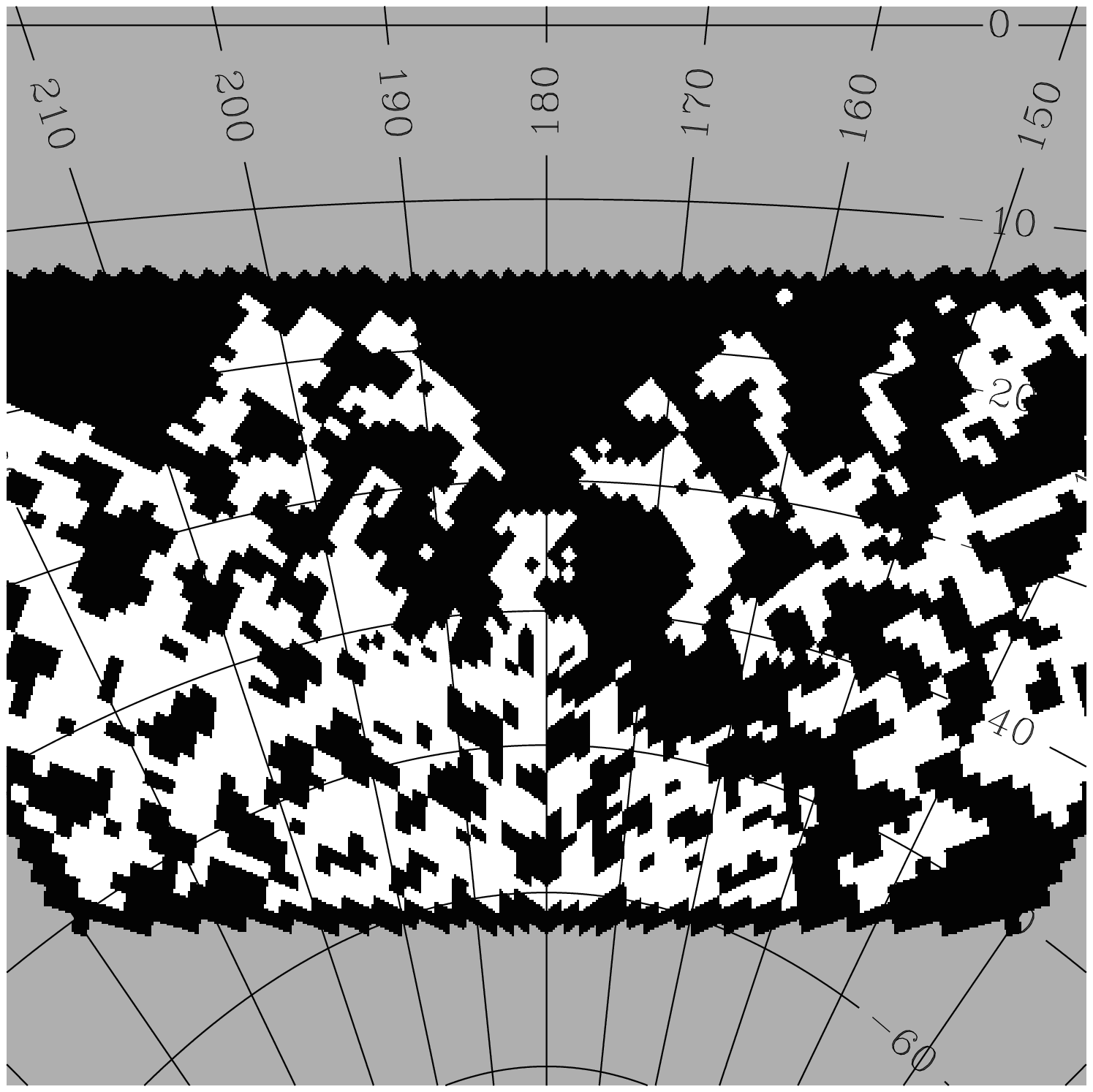}
\includegraphics{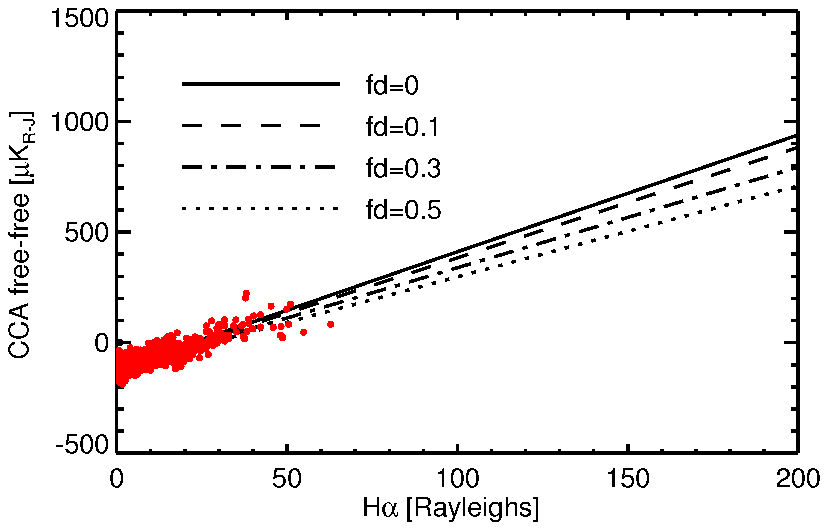}
\caption{T-T analysis for estimation of $T_{\rm e}$. {\it Top}: gnomonic projection showing the mask used (masked pixels are in black, while pixels used in the analysis are in white). {\it Bottom}: T-T plot comparing the CCA free-free solution with the H$\alpha$ template for $f_{\rm d}=0.3$ (points) and linear fits to the T-T plots for different values of $f_{\rm d}$ (lines).}
\label{fig:TT_analysis}
\end{center}
\end{figure}

The ratio $T_{\rm ff}(\nu)/\rm{H}{\alpha}$ can be obtained by comparing the H${\alpha}$ and free-free emission from component separation through a temperature-temperature plot (T-T analysis). We made free-free versus H${\alpha}$  plots by using the CCA free-free solution at 30\,GHz and both \cite{cliveff} and \cite{fink2003} H${\alpha}$ templates corrected for dust absorption for different values of $f_{\rm d}$.  We considered $3^\circ$ resolution maps, sampled with $N_{\rm{side}}=64$ pixels. Besides point sources, we excluded from the analysis the region most affected by dust absorption based on the \cite{schlegel} $E(B-V)$ map, as shown in the top panel of Fig.~\ref{fig:TT_analysis}. 
The electron temperature $T_{\rm e}$  has been inferred by fitting the data points with a linear relation and converting the best-fit slope to $T_{\rm e}$ through Eq.~(\ref{ff2halpha}). The error on $T_{\rm e}$ has been derived from the error on the best-fit slope given by the fitting procedure, through error propagation. 
In the bottom panel of Fig.~\ref{fig:TT_analysis} we show the T-T plots for the \cite{cliveff} H$\alpha$ template corrected for $f_{\rm d}=0.3$ (red points), and the best-fit linear relations to the T-T plots for diffeent values of  $f_{\rm d}=0.3$ (lines). 
The electron temperatures are reported in the top part of Table~\ref{tab:tt_te}.
We obtain $T_{\rm e}=5900$--3900\,K with $f_{\rm d}=0$--0.5 for the Dickinson template; the Finkbeiner template yields generally higher, but consistent, values ($T_{\rm e}=5800$--4300\,K with $f_{\rm d}=0$--0.5). 

\begin{table*}[tmb]                 % table* is a two-column table.  Drop the * for one column.
\begingroup
\newdimen\tblskip \tblskip=5pt
%\caption{}                          % Caption goes here.
\caption{Inferred $T_{\rm e}$ [K] for T-T analysis and C-C analysis using different H${\alpha}$ templates and dust absorption fractions $f_{\rm d}$.}
%\label{}                            % Label goes here.
\label{tab:tt_te}
\nointerlineskip
\vskip -3mm
\footnotesize
\setbox\tablebox=\vbox{
   \newdimen\digitwidth 
   \setbox0=\hbox{\rm 0} 
   \digitwidth=\wd0 
   \catcode`*=\active 
   \def*{\kern\digitwidth}
%
%   \newdimen\signwidth 
   \setbox0=\hbox{+} 
   \signwidth=\wd0 
   \catcode`!=\active 
   \def!{\kern\signwidth}

{\tabskip=2em
%\halign{#\hfil&#\hfil&#\hfil&#\hfil&#\hfil\cr
\halign{#\hfil&#\hfil&#\hfil&#\hfil&#\hfil&#\hfil\cr
\noalign{\doubleline}
%&&TT analysis&&\cr
%\noalign{\vskip 3pt\hrule\vskip 5pt}
Method&Template&$**f_{\rm d}=0.0$&**$f_{\rm d}=0.1$&**$f_{\rm d}=0.3$& $**f_{\rm d}=0.5$\cr
\noalign{\vskip 3pt\hrule\vskip 5pt}
T-T analysis&Dickinson&5900 $\pm$ 1200 &5400 $\pm$ 1000 &4600 $\pm$ 1200 &3900 $\pm$ 1200\cr
%\noalign{\vskip 3pt\hrule\vskip 5pt}
&Finkbeiner&5800 $\pm$ 1400 & 5400 $\pm$ 1200 &4700 $\pm$ 1200 & 4300 $\pm$ *800\cr
%\noalign{\doubleline}
%&&CC analysis&&\cr
%\noalign{\vskip 3pt\hrule\vskip 5pt}
%CC Template&$f_{\rm d}=0.0$&$f_{\rm d}=0.1$&$f_{\rm d}=0.3$& $f_{\rm d}=0.5$\cr
%\noalign{\vskip 3pt\hrule\vskip 5pt}
C-C analysis&Dickinson&5300 $\pm$ 1500& 4500 $\pm$ 1400&3100 $\pm$ 1100 &2400 $\pm$ 1000\cr
&Finkbeiner&7000 $\pm$ 1700& 6500 $\pm$ 1500&5200 $\pm$  1300 &3800 $\pm$ 1100\cr
\noalign{\vskip 5pt\hrule\vskip 3pt}}}
}
%\endPlancktable                    % ends one-column \halign
\endPlancktablewide                 % ends two-column \halign
%%\tablenote a Footnote a.\par
%%\tablenote b Footnote b.\par
\endgroup
\end{table*}                        % table* is a two-column table.  Drop the * for one column.
%%
%DDD                         5300           4500         3100          2400
%Finkbeiner               7000           6500         5200          3800
\subsection{Comparison with cross-correlation with templates}
An alternative way to compute $T_{\rm ff}(\nu)/\rm{H}{\alpha}$ and $T_{\rm e}$ is through cross-correlation of the H$\alpha$ template with frequency maps (C-C analysis). 
We cross-correlate simultaneously the templates for free-free, dust, and synchrotron emission, as described in \cite{ghosh2011}.
We used the 408\,MHz map from \cite{haslam} as a tracer of synchrotron emission, \cite{cliveff} H$\alpha$ as a tracer of free-free emission and the \cite{finkbeiner1999} model eight 94\,GHz prediction as a tracer of dust emission. We used the same resolution, pixel size and sky mask adopted for the T-T analysis ($3^\circ$ and $N_{\rm side}=64$). As pointed out by \cite{ghosh2011}, at this resolution the template-fitting analysis is more reliable than at $1^\circ$ because the smoothing reduces artifacts in the templates. The correlation coefficients are computed for each emission process at a given frequency by minimizing the generalized $\chi^2$ expression. We also fitted for an additional monopole term that can account for offset contributions in all templates and the data in a way that does not bias the results \citep{macellari2011}. The chance correlation of the templates with the CMB component in the data causes a systematic error in the correlated coefficients and  has been estimated using simulations. We generated 1000 random realizations of the CMB using the {\it WMAP} best-fit $\Lambda$CDM model \footnote{http://lambda.gsfc.nasa.gov/product/map/dr4/pow\_tt\_spec\_get.cfm} and cross-correlated each of them using the templates with the same procedure applied to the data. The amplitude of the predicted chance correlation, given by the RMS over the 1000 realizations, is 1.13\,$\mu$K$_{\rm CMB}/\mu$K$_{\rm CMB}$ for the dust template, 1.12\,$\mu$K$_{\rm CMB}$/R for the free-free template and 3.8\,$\mu$K$_{\rm CMB}$/K for the synchrotron template.

\begin{figure}
%fig8
\begin{center}
\includegraphics[width=88mm]{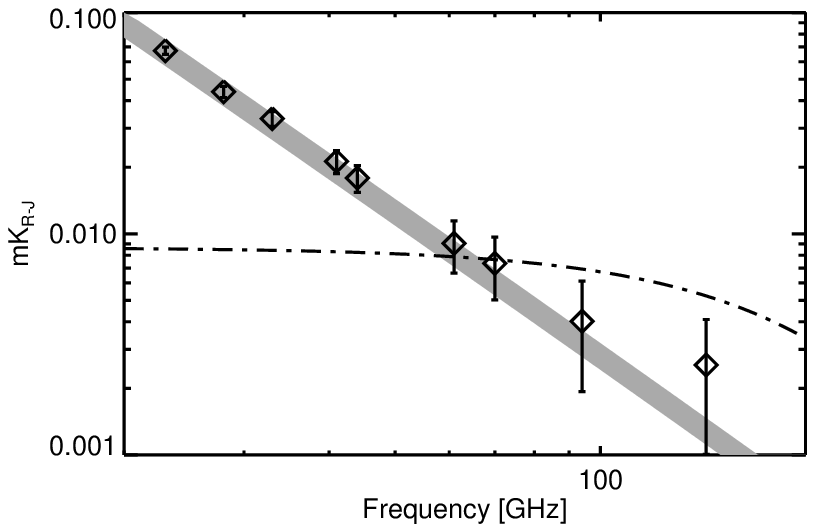}
\includegraphics[width=88mm]{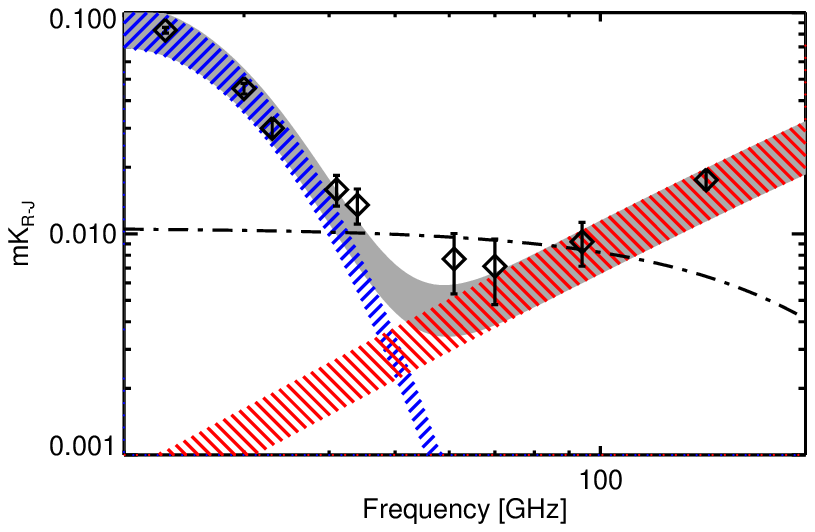}
\caption{Comparison between correlation coefficients (symbols with error bars) and component separation results (shaded areas) for free-free emission ({\it top}) and dust emission ({\it bottom}). For free-free emission we show the \cite{cliveff} H$\alpha$ correlation coefficients and for dust emission the \cite{finkbeiner1999} correlation coefficients. The grey area in the bottom panel is the sum of the AME (blue) and thermal dust (red) components. The dash-dotted line in both panels shows the 1$\sigma$ error due to the chance correlation of CMB with foreground templates, estimated using simulations. \label{fig:anna_tuhin}}
\end{center}
\end{figure}

In the top panel of Fig.~\ref{fig:anna_tuhin} we compare the H$\alpha$ correlation coefficients (points with error bars) with the component separation results for free-free emission obtained in Sect.~\ref{sec:GLS} (shaded area). The flux for both the component separation and cross-correlation has been computed as the standard deviation of the maps (the separated free-free map and the scaled H$\alpha$ template, respectively) as this is not affected by possible offsets between the \Planck~data and the H$\alpha$ template. 
There is generally good agreement between the two results; in the frequency range 40--60\,GHz there is an excess in the correlated coefficients, which could be indicative of a contribution from the AME component \citep[similar to that found by][]{dob1}. Flattening of the C-C coefficients for $\nu>60$\, GHz is consistent with positive chance correlation between the CMB and the H$\alpha$ template. 

The dust-correlated coefficients are compared with the component separation results in the bottom panel of Fig.~\ref{fig:anna_tuhin}. The agreement is very good for $\nu<40$\,GHz and $\nu>100$\,GHz, where AME and thermal dust emission are strong. In the 40--70\,GHz range the C-C results are higher than the component separation results. As discussed in Appendix~\ref{sec:appa}, the parametric fit to the AME spectrum implemented by CCA could be inaccurate in this frequency range, where the AME is faint. Alternatively, a similar effect could be explained by the presence of a secondary AME peak, around 40\,GHz (e.g. \citealt{planck2011-7.2}, \citealt{ghosh2011}) or flattening of the dust spectral index towards low frequencies, which are not included in our spectral model.  Discriminating between these hypotheses is not possible given the large error bars. 

To determine the free-free electron temperature the H$\alpha$ correlation coefficients have been fitted with a combination of power-law free-free radiation (with fixed spectral index of $-2.14$) and a CMB chance correlation
term (which is constant in thermodynamic units). The amplitude of the free-free component with respect to H$\alpha$ resulting from the fit, and its uncertainty, yield $T_{\rm e}$ and the corresponding error bar.
The results for the Gould Belt region outside the adopted sky mask are reported in the bottom part of Table~\ref{tab:tt_te}. We find $T_{\rm e}=5300$--2400\,K for $f_{\rm d}=0$--0.5 with the \cite{cliveff} template, and  $T_{\rm e}=7000$--3800\,K for $f_{\rm d}=0$--0.5 with the \cite{fink2003} template.  

With respect to the T-T analysis, these results are more sensitive to the choice of the template and the $f_{\rm d}$ correction. Similarly, we espect the C-C analysis to be more sensitive to the other systematic uncertainties on the templates, such as the contribution of scattered light to the H$\alpha$ map (\citealt{witt}, \citealt{brandt2012}). 

The sensitivity of the C-C analysis to differences between the Dickinson et al. and Finkbeiner templates --- the former yielding lower $T_{\rm e}$ than the latter --- is a known issue (see \citealt{ghosh2011} for a detailed analysis). The different processing of the two maps results in residuals at the 1\,R level over large regions of the sky, and of more than 20\,R near very bright regions. The adopted $\chi^2$ estimator, which contains the square of the template in the denominator, tends to amplify the differences. 

For this analysis we adopted a 3$^\circ$ resolution, as advised by \cite{ghosh2011} to reduce artefacts in the templates due to beam effects, and we masked the most discrepant pixels. Still, the best-fit electron temperatures yielded by the two templates may differ by 30\,\%, whereas for the T-T analysis this difference is 10\% at most. In fact, the fit of the T-T plot is determined by large samples of pixels, on which the two templates are generally more similar, while the  C-C method is more sensitive to bright features, on which they may be more different. We verified that, by enlarging the mask to exclude the brightest pixels, the numbers we obtain for the two templates get in better agreement. 

The C-C results are always consistent with the T-T ones within the error bars; however, we note that, for the Dickinson et al. template, they are systematically lower. 
Besides systematic errors related to methods and templates, a difference between T-T and C-C results could also indicate spatial variability of $T_{\rm e}$ within the region, since the two methods have different sensitivity to different features in the map. This confirms that estimating the free-free electron temperature is a difficult problem and that caution is needed when interpreting the results.

\section{AME as spinning dust emission.}\label{sec:ame}
An explanation that is often invoked for the AME is electric dipole radiation from small, rapidly spinning, Polycyclic Aromatic Hydrocarbon (PAHs) dust grains (\citealt{erickson1957}, \citealt{DL1998}, \citealt{dob1}, \citealt{dob2}, \citealt{dob4}). 

Alternatively, the AME could be due to synchrotron radiation with a flat (hard) spectral index \citep[e.g.][]{bennett2003a}. 
The presence of such a hard spectrum synchrotron component could be highlighted by comparing the 408\,MHz map of \cite{haslam}, which would predominantly trace steep spectrum radiation, with the 2.3\,GHz map by \cite{jonas}, which would be more sensitive to flat spectrum radiation. This issue has been studied in detail by \cite{peel2011} using a cross-correlation of {\it WMAP} 7-yr data with foreground templates. They analysed the region defined by $170^\circ \leq l \leq 210^\circ$, $-55^\circ \leq b \leq -25^\circ$ and found that the dust-correlated coefficients are mostly unaffected by the use of the 2.3\,GHz template instead of the 408\,MHz template. This indicates that hard synchrotron radiation cannot account for most of the dust-correlated component at low frequencies.

To check the hypothesis of spinning dust emission we applied the method proposed by \cite{ysard2011}, which exploits the {\tt SpDust} (\citealt{spdust1}, \citealt{spdust2}) and {\tt DustEM} \citep{compiegne2011} codes, to model the frequency spectra of thermal and anomalous dust emission from the microwaves to the IR.
The dust populations and properties are assumed to be the same as in the diffuse interstellar medium at high Galactic latitude (DHGL), defined in \cite{compiegne2010}. This model includes three dust populations: PAHs; amorphous carbonaceous grains; and amorphous silicates. For PAHs, it assumes a log-normal size distribution with centroid $a_0 = 0.64$\,nm and width $\sigma = 0.4$, with a dust-to-gas mass ratio $M_{\rm PAH}/M_{\rm H} = 7.8\times 10^{-4}$. 

By fitting the thermal dust spectrum with {\tt DustEM} we determine the local intensity of the interstellar radiation field, $G_0$ (the scaling factor with respect to a UV flux of $1.6 \times 10^{-3}$\,erg\,s$^{-1}$cm$^{-2}$ integrated between 6 and 13.6\,eV), and the hydrogen column density, $N_{\rm H}$. We then fit the AME spectrum with {\tt SpDust}, the only free parameter being the local hydrogen density $n_{\rm H}$. We assume a cosmic-ray ionization rate $\zeta_{\rm CR} = 5\times 10^{-17}$\,s$^{-1}$H$^{-1}$, and take the electric dipole moment to be as in \cite{DL1998}, a prescription also shown to be compatible with the AME extracted from {\it WMAP} data \citep{ysard2010}. 
It is worth noticing that there is a degeneracy with the size of the grains (smaller size yields higher peak frequency and intensity of the AME). However, the size distribution can only be constrained using shorter wavelength data (typically 3--8$\microns$). The size we are adopting (0.64\,nm) is motivated by its ability to reproduce the data in the mid-IR \citep{compiegne2011}; other models adopt different sizes (e.g. 0.54\,nm and 0.5\,nm in \citealt{Dli2001} and \citealt{DLi2007} respectively). 

\begin{figure}
%fig9
\begin{center}
\includegraphics[width=88mm]{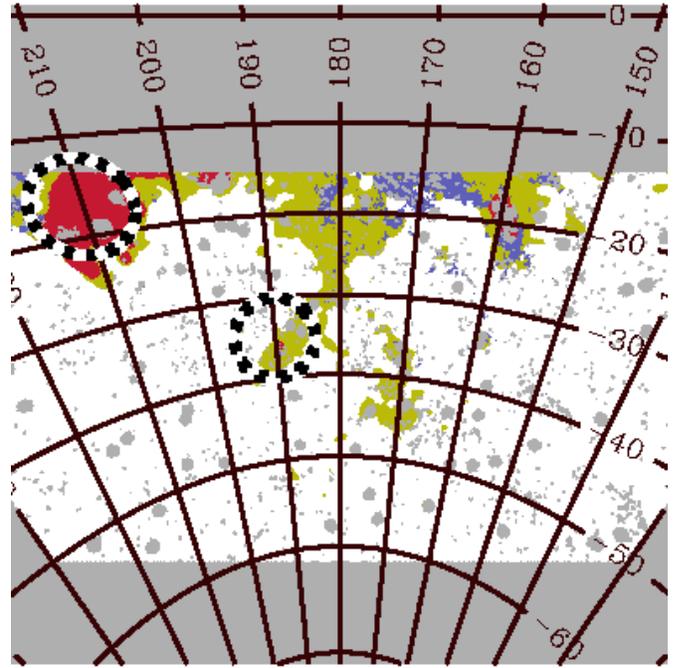}
\caption{Partitioning of the Gould Belt region based on thresholds over free-free emission (red), CO emission (blue), and total emission at 30\,GHz (yellow), used as tracers of \ion{H}{ii} gas, molecular gas, and diffuse ISM environments, respectively, and the rest of the Gould Belt region (light blue). Circled regions are those selected for the computation of spectra and have been labelled as \ion{H}{ii}-gas (Barnard's arc) and diffuse ISM (centred on $l=190^\circ$, $b=-35^\circ$) regions, respectively. \label{fig:environments_map}}
\end{center}
\end{figure}

As the Gould Belt region contains strong foreground emission components,  significantly correlated with each-other, we expect different environments to be mixed in a complex way. In order to obtain meaningful results for the physical modelling we tried to isolate sub-regions where single environments dominate.
To first order, we can use the free-free emission as a tracer of the ionized gas environment, CO emission as a tracer of molecular gas, and associate the rest of the emission with the diffuse ISM.
In Fig.~\ref{fig:environments_map} we schematically map the different environments by setting a threshold on the free-free emission coming from component separation, the CO emission from \Planck, and the total foreground emission at 30\,GHz. We identified two relatively big sub-regions (shown as circles in Fig.~\ref{fig:environments_map}) as selections which are dominated by ionized gas and diffuse ISM environments. It would not be meaningful to consider smaller areas because of the patch-by-patch estimation of the AME frequency scalings, which means that our AME spectra are averaged over relatively large areas of the sky. Due to the clumpiness of the molecular gas environment it was not possible to select a region for this case. It is worth noting that some molecular gas may be contained in the diffuse ISM region. 

The spectra of AME and thermal dust in the 20--353\,GHz frequency range are based on component separation results. The frequency scaling is that estimated with CCA and the normalization is given by the average of the reconstructed amplitude map in the region of the sky considered. The error bars on the data points include the RMS of the amplitude in the same region (considered as the error in the normalization) and the errors on the estimated spectral parameters. The thermal dust spectra have been complemented with higher frequency data points computed directly from the frequency maps: \Planck~545\,GHz and 857\,GHz; IRIS 100$\microns$ map; and the IRIS 12$\microns$ map corrected for Zodiacal light emission used in \cite{ysard2010}.

\begin{figure}
%fig10
\begin{center}
\includegraphics[width=88mm]{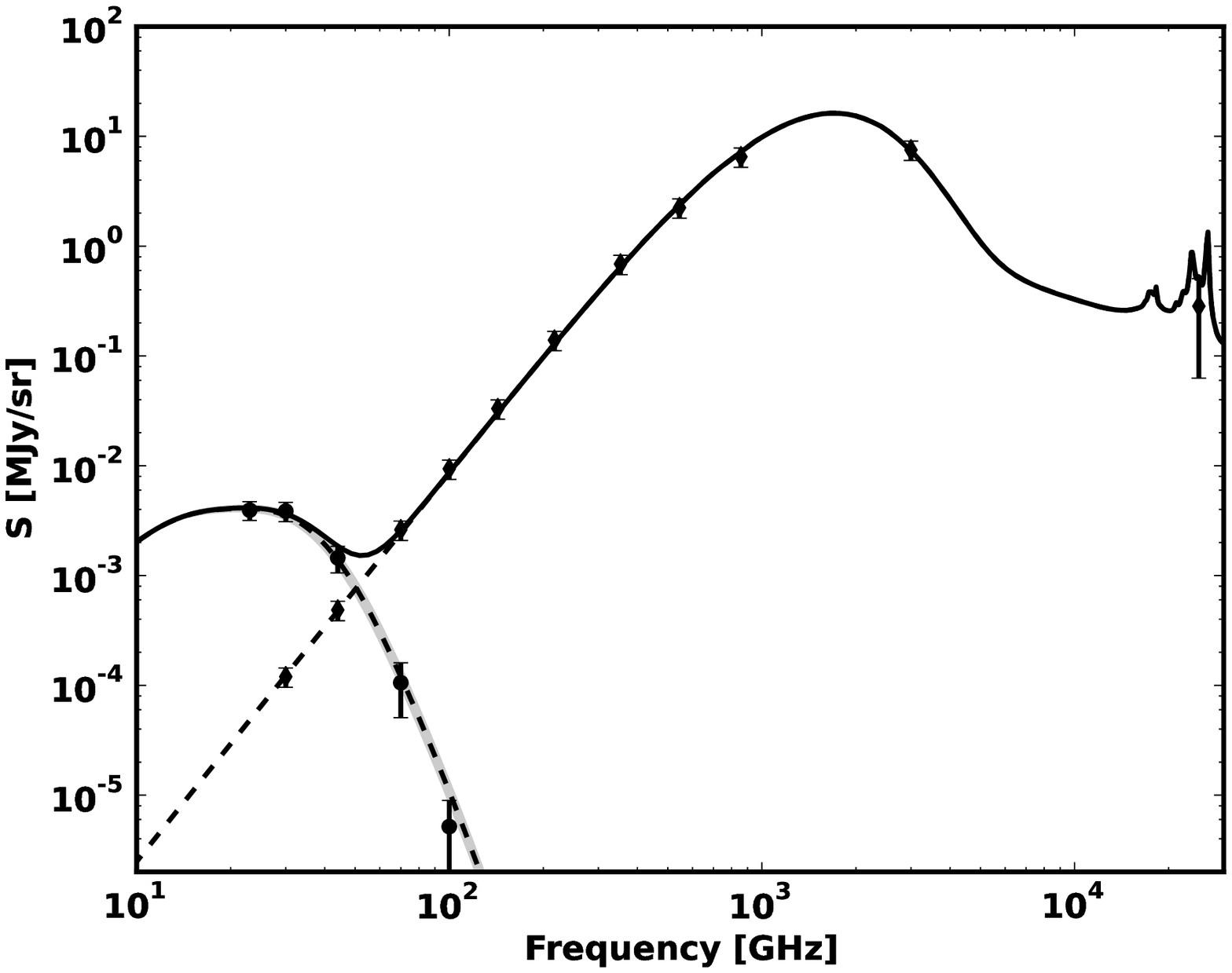}
\includegraphics[width=88mm]{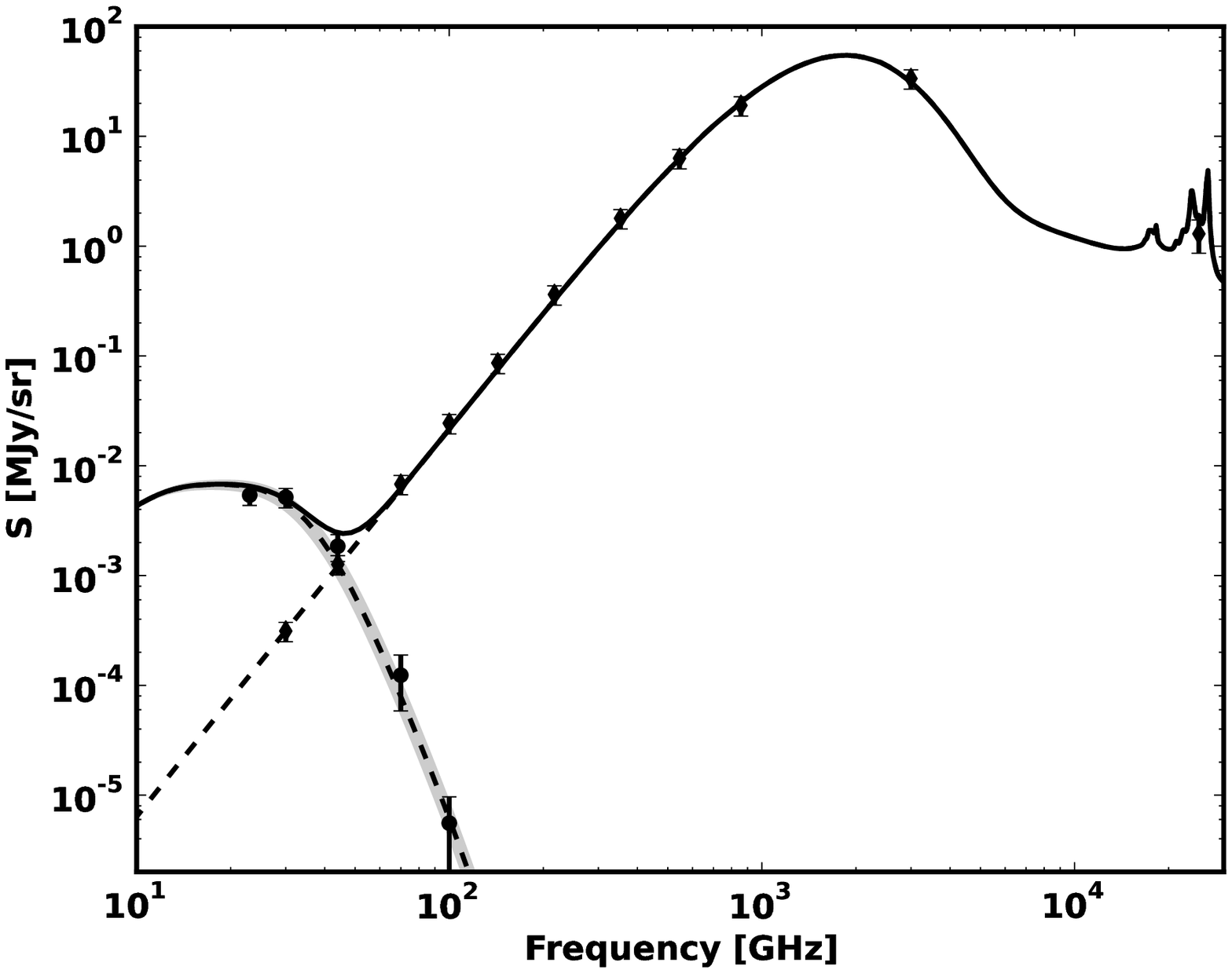}
\includegraphics[width=88mm]{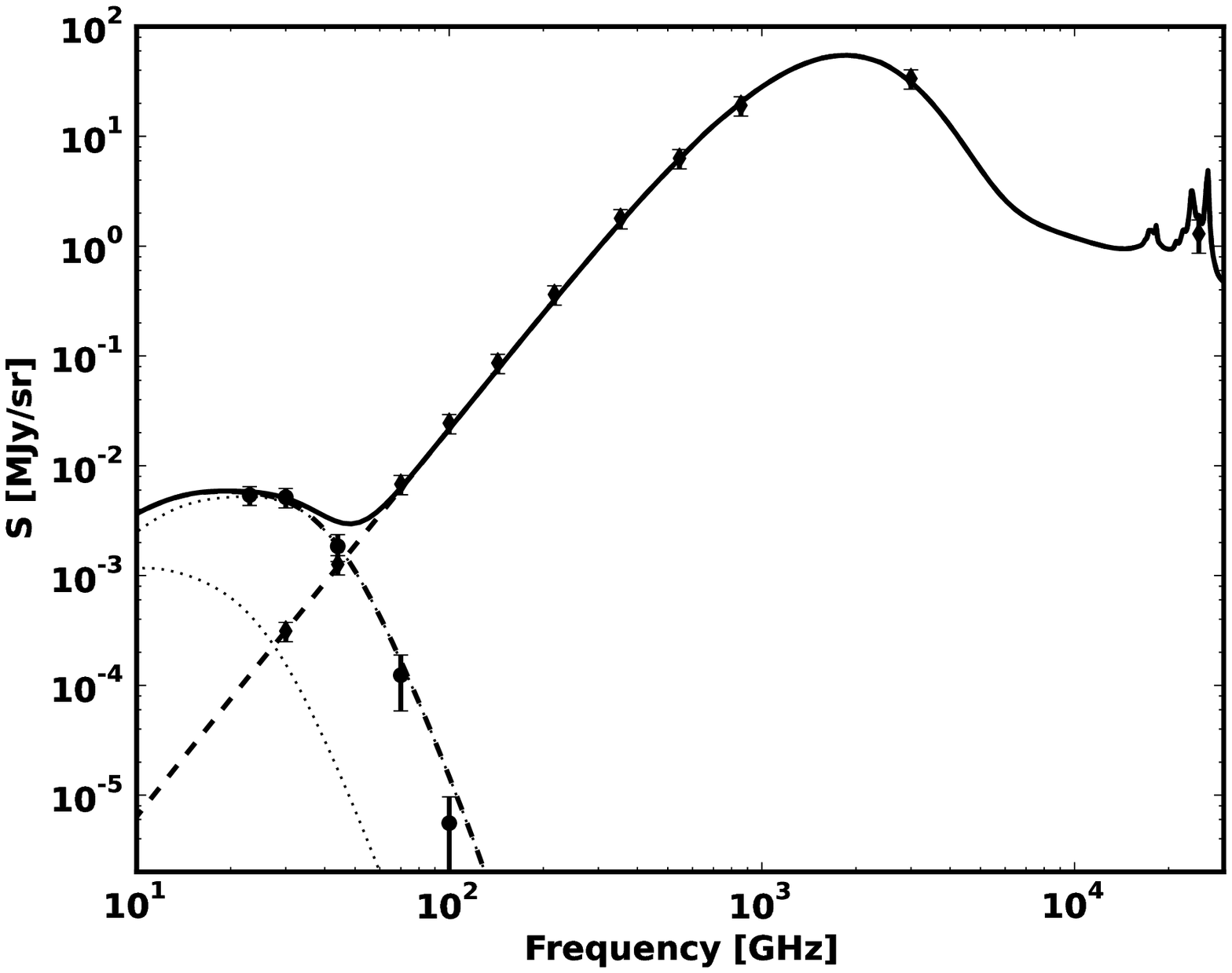}
\caption{Frequency spectra (black points with error bars) for thermal dust emission and AME compared, respectively, with {\tt DustEM} and {\tt SpDust} (dashed lines) for the diffuse ISM ({\it top}) and ionized gas ({\it middle and bottom}) regions within the Gould Belt. The solid line is the sum of the {\tt DustEM} and {\tt SpDust} models. The grey area on the top and middle panels correspond to the $\pm 1\,\sigma$ variations on the best-fit $n_{\rm H}$ values, when fitting for a single phase. In the bottom panel we consider a mixture of two phases ($n_{\rm H} = 0.1$\,cm$^{-3}$ and $n_{\rm H} = 55$\,cm$^{-3}$, in the proportion of 46\,\% and 54\,\%, respectively), which marginally improves the fit for the ionized gas region at 23\,GHz (the error is 0.3\,$\sigma$ instead of 0.9\,$\sigma$ ).}
\label{fig:nathalie}
\end{center}
\end{figure}

The results of the modelling for the ionized gas and diffuse ISM regions within the Gould Belt are shown in Fig.~\ref{fig:nathalie}. 
The empirical spectra of AME coming from component separation can be successfully modelled as spinning dust emission for both regions.
The match between data and model becomes worse at higher frequencies, where the AME spectrum could be biased (see Sect.~\ref{sec:simulations} and Appendix~\ref{sec:appa}). 

The joint fit of thermal and spinning dust models yields plausible physical descriptions of the two environments. In the top panel of Fig.~\ref{fig:nathalie} the diffuse ISM region is modelled with $N_{\rm H} = 2.46\times10^{21}$\,H\,cm$^{-2}$, $G_0 = 0.55$ and $n_{\rm H}$=50\,cm$^{-3}$. The ionized region (middle panel) is modelled with $N_{\rm H} = 5.73 \times10^{21}$ H\,cm$^{-2}$, $G_0 = 0.90$ and  $n_{\rm H} =$25\,cm$^{-3}$. 

We tested the stability of these results against calibration errors on the high frequency \Planck~(545 and 857\,GHz) and IRIS (100$\microns$ and 12$\microns$) data (the remaining data points come from the component separation procedure and their error bars already include systematic uncertainties). 

The total calibration uncertainty on the \Planck~545 and 857\,GHz channels is estimated to be 10\,\% \citep{planck2013-p03b}; that on the IRIS data is of the order 10\% or larger, especially at 12$\microns$ where it also includes errors on the zodiacal light subtraction. We have verified that very conservative uncertainties up to 20\,\% both on \Planck~ and IRIS data have negligible impact on $G_0$, while they may affect $N_{\rm H}$ and $n_{\rm H}$ (up to a level of about 10\,\%). 
The overall picture however does not change: the ionized region is less dense and illuminated by a stronger radiation field than the diffuse 
region (which is expected to contain mostly neutral gas). 
Both the spectra can be modelled as spinning dust emission arising from 
regions with densities characteristics of the cold neutral medium 
(CNM, a few tens of H per cm$^3$). This confirms the results 
of \cite{planck2011-7.2} And \cite{planck2011-7.3}, showing 
that most of the observed AME could be explained by spinning 
dust in dense gas. In fact, whenever we have a mixture of warm neutral medium (WNM), warm ionized medium (WIM) and CNM, the spinning dust
spectrum is dominated by the denser phase, which emits more strongly.  

In the bottom panel of Fig.~\ref{fig:nathalie} we consider for the ionized region a mixture of two phases, one having lower density ($n_{\rm H} = 0.1$\,cm$^{-3}$, 46\,\%) and one having higher density ($n_{\rm H} = 55$\,cm$^{-3}$, 54\,\%), illuminated by the same $G_0$ as in the middle panel. 
Such a mixture fits the data somewhat better at 23\,GHz than the one-phase model considered previously (the error in the fit at this frequency being 0.3\,$\sigma$ instead of 0.9\,$\sigma$). In order to fully isolate and study different 
ISM phases  (ionized/neutral, dense/diffuse), both the observations and 
the analysis should be carried out at high angular resolution. 

\section{Conclusions} \label{sec:conclu}
We performed an analysis of the diffuse low-frequency Galactic foregrounds as seen by \Planck~in the Southern part ($130^\circ\leq l\leq 230^\circ$ and $-50^\circ\leq b\leq -10^\circ$) of the Gould Belt system, a local star-forming region emitting bright diffuse foreground emission. Besides \Planck~data our analysis includes {\it WMAP} 7-yr data and foreground ancillary data as specified in Table~2. 

We used the CCA (\citealt{bonaldi2006}, \citealt{ricciardi2010}) component separation method to disentangle the diffuse Galactic foregrounds.
In the region of interest the synchrotron component is smooth and faint. 

The free-free emission is strong and it clearly dominates in the Orion-Barnard region. We inferred the free-free electron temperature both by cross-correlation (C-C) of channel maps with foreground templates and temperature-temperature (T-T) plots comparing the CCA free-free emission with H$\alpha$ maps. We obtained $T_{\rm e}$ ranging from 3100 to 5200\,K for $f_{\rm d}$=0.3, which broadens to 2400--7000\,K when we allow $f_{\rm d}$ to range within 0--0.5.  
The use of the \cite{fink2003} H$\alpha$ template yields systematically higher $T_{\rm e}$ than the \cite{cliveff} one. In the case of the T-T analysis the difference is at most 500\,K ($<1\sigma$), while for the C-C analysis it can reach 2000\,K (within $2\,\sigma$). The C-C results for the Dickinson et al. template are also systematically lower than the T-T ones, yet consistent within 1$\sigma$.

The AME is the dominant foreground emission at the lowest frequencies of \Planck~over most of the region considered. 
We estimated the AME peak frequency in flux density units to be $25.5\pm 1.5$\,GHz, almost uniformly over the region of interest. This is in agreement with AME spectra measured in compact dust clouds \citep[e.g.][]{planck2011-7.2} and {\it WMAP} 9-yr results at low latitudes (once the same convention is adopted, e.g. their AME spectrum is converted from K$_{\rm R-J}$ to flux density, \citealt{bennett2012}). In the case of diffuse AME at higher latitudes a lower peak frequency is favoured (\citealt{banday2003}, \citealt{davies2006}, \citealt{ghosh2011}, \citealt{special}). Spatial variability of the peak frequency of AME is expected, in the case of spinning dust emission, as a result of changes in the local physical conditions. For instance, the observed differences can be modelled in terms of a different density of the medium (lower density at high latitudes causes lower peak frequency) or a different size of the grains (smaller size giving higher peak frequency). The ability of our method to correctly recover the peak frequency of the AME, $\nu_{\rm p}$, has been verified through realistic simulations. We also considered the effect of systematic errors in the spectral model and in the free-free template and we demonstrated that they have negligible impact on $\nu_{\rm p}$. 

Following \cite{peel2011}, a hard (flat spectrum) synchrotron component would not be sufficient to account for the dust-correlated low-frequency emission in this region. In support of the spinning dust mechanism, we performed a joint modelling of vibrational and rotational emission from dust grains as described by \cite{ysard2011} and we obtained a good description of the data from microwaves to the IR. The fit, which we performed separately for the ionized area near to Barnard's arc and the diffuse emission towards the centre of our region, yields in both cases plausible values for the local density and radiation field. This indicates that the spinning dust mechanism can reasonably explain the AME in the Gould Belt.

\section{Acknowledgements}
Based on observations obtained with \Planck~(\url{http://www.esa.int/Planck}), an ESA science mission with instruments and contributions directly funded by ESA Member States, NASA, and Canada.

The development of Planck has been supported by: ESA; CNES and CNRS/INSU-IN2P3-INP (France); ASI, CNR, and INAF (Italy); NASA and DoE (USA); STFC and UKSA (UK); CSIC, MICINN, JA and RES (Spain); Tekes, AoF and CSC (Finland); DLR and MPG (Germany); CSA (Canada); DTU Space (Denmark); SER/SSO (Switzerland); RCN (Norway); SFI (Ireland); FCT/MCTES (Portugal); and The development of Planck has been supported by: ESA; CNES and CNRS/INSU-IN2P3-INP (France); ASI, CNR, and INAF (Italy); NASA and DoE (USA); STFC and UKSA (UK); CSIC, MICINN and JA (Spain); Tekes, AoF and CSC (Finland); DLR and MPG (Germany); CSA (Canada); DTU Space (Denmark); SER/SSO (Switzerland); RCN (Norway); SFI (Ireland); FCT/MCTES (Portugal); and PRACE (EU).

A description of the Planck Collaboration and a list of its members, including the technical or scientific activities in which they have been involved, can be found at \url{http://www.sciops.esa.int/index.php} \url{?project=planck&page=Planck_Collaboration}. We acknowledge the use of the HEALPix \citep{gorski2005} package and of the LAMBDA website \url{http://lambda.gsfc.nasa.gov}.
\bibliography{biblio}

\newpage
\appendix
\section{Harmonic-domain CCA}\label{sec:method}
The sky radiation, $\tilde{x}$, from direction ${r}$ at frequency $\nu$ results from the superposition of signals coming from $N_{\rm c}$ different physical processes $\tilde{s}_j$:
\begin{equation}
\tilde{x}({r},\nu)=\sum_{j=1}^{N_{\rm c}}\tilde{s}_j({r},\nu).
\end{equation}
The signal $\tilde{x}$ is observed through a telescope, the beam pattern of which can be modelled, at each frequency, as a spatially invariant point spread function $B({r},\nu)$.  For each value of $\nu$, the telescope convolves the physical radiation map with $B$. The frequency-dependent convolved
signal is input to an $N_{\rm d}$-channel measuring instrument, which
integrates the signal over frequency for each of its channels and adds
noise to its outputs.  The output of the measurement channel at a generic frequency $\nu$ is
\begin{equation}
x_\nu({r})=\int B({r}-{r'},\nu')\sum_{j=1}^{N_{\rm c}}t_\nu(\nu')
\tilde{s}_j({r',\nu'}) dr' d\nu'+n_\nu({r})\label{m0},
\end{equation}
where $t_\nu(\nu')$ is the frequency response of the channel and
$n_\nu({r})$ is the noise map.  The data model
in Eq.~(\ref{m0}) can be simplified by virtue of the following
assumptions:
\begin{itemize}
\item Each source signal is a separable function of direction and frequency, i.e.,
\begin{equation}
\tilde{s}_j({r},\nu)={s}_j({r})f_j(\nu)\label{3};
\end{equation}
\item $B({r},\nu)=B_\nu({r})$ is constant within the bandpass of the
measurement channel.
\end{itemize}
These two assumptions lead us to a new data model:
\begin{equation}
x_\nu({r})=B_\nu({r})*\sum_{j=1}^{N_{\rm c}}h_{\nu j}{s_j}({r}) +n_\nu({r})\label{m1},
\end{equation}
where $*$ denotes convolution, and
\begin{equation}
h_{\nu j}\equiv \int t_\nu(\nu')f_j(\nu')d\nu' \label{mixmat}.
\end{equation}
For each location, $r$, we define:
\begin{itemize}
\item the $N_{\rm c}$-vector $\vec{{s}}$ (sources vector) whose elements are ${s_j}({r})$;
\item the $N_{\rm d}$-vector $\vec{{x}}$ (data vector) whose elements are ${x_\nu}({r})$;
\item the $N_{\rm d}$-vector $\vec{{n}}$ (noise vector) whose elements are $n_\nu({r})$;
\item the diagonal $N_{\rm d}$-matrix $\tens{B}$ whose elements are $B_\nu({r})$;
\item the $N_{\rm d}\times N_{\rm c}$ matrix $\tens{H}$ containing all $h_{\nu j}$ elements.
\end{itemize}
Then, we can rewrite Eq.~(\ref{m1}) in vector form:
\begin{equation}
\vec{x}(r)=[\tens{B}*\tens{H}\vec{s}](r)+\vec{n}(r).\label{vect_m1}
\end{equation}
The matrix $\tens{H}$ is called the mixing matrix and contains the frequency scaling of the components for all the data maps involved. 

When working in the pixel domain, under the assumption that $\tens{B}$ does not depend on the frequency, we can simplify Eq.~(\ref{vect_m1}) to
\begin{equation}
\vec{x}=\tens{H}\vec{s}+\vec{n},
\end{equation}
where the components in the source vector $\vec{s}$ are now convolved with the instrumental beam. 

Eq.~(\ref{vect_m1}) can be translated to the harmonic domain,
where, for each transformed mode, it becomes
\begin{equation}
\vec{X}=\widetilde{\tens{B}}\tens{H}\vec{S}+\vec{N}\label{modhcca},
\end{equation}
where $\vec{X}$, $\vec{S}$, and $\vec{N}$ are the transforms
of $\vec{x}$, $\vec{s}$, and $\vec{n}$, respectively,
and $\widetilde{\tens{B}}$ is the transform of matrix $\tens{B}$.
Relying on this  data model we can derive the following relation between  the cross-spectra of the data $\widetilde{\tens{C}}_{\vec{x}}(\ell)$, sources $\widetilde{\tens{C}}_{\vec{s}}(\ell)$ and noise, $\widetilde{\tens{C}}_{\vec{n}}(\ell)$, all depending on the multipole $\ell$:
\begin{equation}
\widetilde{\tens{C}}_{\vec{x}}(\ell)=\widetilde{\tens{B}}(\ell)\tens{H}\widetilde{\tens{C}}_{\vec{s}}(\ell)\tens{H}^{\rm T}\widetilde{\tens{B}}^\dagger(\ell)+\widetilde{\tens{C}}_{\vec{n}}(\ell),
\label{hcca_constr}
\end{equation}
where the dagger superscript denotes
the adjoint matrix. To reduce the number of unknowns, the mixing matrix is parametrized through a parameter vector $\vec{p}$ (such that
$\tens{H}=\tens{H}(\vec{p})$), using the fact
that its elements are proportional to the spectra of astrophysical
sources (see Sect.~\ref{sec:datamodel}). 

%{\bf} 

Since the foreground properties are expected to be spatially variable, we work on relatively small square patches of data. This allows us to use the 2D Fourier transform to approximate the harmonic spectra \citep[see, e.g.,][]{bondefstathiou1987}.

The HEALPix \citep{gorski2005} data on the sphere are projected on the plane tangential to the centre of the patch and re-gridded with a suitable number of bins in order to correctly sample the original resolution. Each pixel in the projected image is associated with a specific vector normal to the tangential plane and it assumes the value of the HEALPix pixel nearest to the corresponding position on the sphere. Clearly, the projection and re-gridding process will create some distortion in the image at small scales and will modify the noise properties. However, we verified that this has negligible impact on the spectra in Eq. ~(\ref{hcca_constr}) for the scales considered in this work and, therefore, on the spectral parameters.
If $\vec{x}(i,j)$ contains the data projected on the planar grid and $\vec{X}(i,j)$ is its  2-dimensional discrete Fourier transform, 
the energy of the signal at a certain scale, which corresponds to the power spectrum, can be obtained as the average of $\vec{X}(i,j) \vec{X}^{\dag}(i,j)$ over annular bins $D_{\hat{\ell}}$, $\hat{\ell}=1,\ldots,\hat{\ell}_{\rm max}$ \citep{CCA_fourier}:
\begin{equation}
\widetilde{\tens{C}}_{\vec{x}}(\hat{\ell}) = \frac{1}{M_{\hat{\ell}}}
\sum_{i,j 
\in D_{\hat{\ell}}} \vec{X}(i,j)  \vec{X}^{\dag}(i,j), \label{dataspectrum}
\end{equation}
where $M_{\hat{\ell}}$ is the number of pairs $(i,j)$ contained in the spectral  bin denoted by $D_{\hat{\ell}}$. Every spectral bin $\hat \ell$ is related to a specific $\ell$ in the spherical harmonic domain by
\begin{equation}
\ell=(\hat \ell-1)\,2p \Delta_\ell/N_{\rm pix}
\end{equation}
where $p$ is the thickness of the annular bin, $\Delta_\ell=180/L_{\rm deg}\,(N_{\rm pix}-1)$, and $L_{\rm deg}$, $N_{\rm pix}$ are the size in degrees and the number of pixels on the side of the square patch, respectively.

If we reorder the matrices 
${\tens{C}}_{\vec{x}}(\hat\ell) - 
{\tens{C}}_{\vec{n}}(\hat \ell)$ and 
${\tens{C}}_{\vec{s}}(\hat \ell)$ into vectors 
$\vec{d}(\hat \ell)$ and $\vec{c}(\hat \ell)$, respectively, we can 
rewrite Eq.~(\ref{hcca_constr}) as 
\begin{equation}
\vec{d}(\hat{\ell}) = \tens{H}_{k}(\hat{\ell})\vec{c}(\hat{\ell})+ \vec{\epsilon}(\hat{\ell}),\label{fd_cca_error}
\end{equation}
where $\tens{H}_{k}(\hat \ell) = 
[\widetilde{\tens{B}}(\hat \ell)\tens{H}]\otimes[\widetilde{\tens{B}}(\hat \ell)\tens{H}]$, and the symbol $\otimes$ denotes the Kronecker product. The vector $\vec{d}(\hat{\ell})$ is now computed using the approximated data cross-spectrum matrix in Eq.~(\ref{dataspectrum}) and $\vec{\epsilon}(\hat{\ell)}$ represents the error on the noise power spectrum.

The parameter vector $\vec{p}$ and the source cross-spectra are finally obtained by minimizing the functional:
\begin{eqnarray}\label{hcca_objective}
\!\!&&\vec{\Phi}[\vec{p},\vec{c}_V]=\\
\!\!&&\!\!\!\![\vec{d}_V-\tens{H}_{kB}(\vec{p})\cdot \vec{c}_V]^T \tens{N}_{\epsilon B}^{-1} [\vec{d}_V-\tens{H}_{kB}(\vec{p})\cdot \vec{c}_V]+\lambda \vec{c}_V^{\rm T}\tens{C}\vec{c}_V. \nonumber
\end{eqnarray}
The vectors $\vec{d}_V$ and $\vec{c}_V$ contain the elements $\vec{d}(\hat \ell)$ and $\vec{c}(\hat \ell)$, respectively, and the diagonal matrices $\tens{H}_{kB}$ and $\tens{N}_{\epsilon}$ the elements  $\tens{H}_{k}(\hat \ell)$ and the covariance of error $\vec{\epsilon}(\hat \ell)$ for all the relevant spectral bins.
The term $\lambda\vec{c}_V^{\rm T}\tens{C}\vec{c}_V$ is a
quadratic stabilizer for the source power cross-spectra: the matrix
$\tens{C}$ is in our case the identity matrix, and the parameter $\lambda$ must be tuned to balance the effects of data fit and regularization in the
final solution.
The functional in Eq.~(\ref{hcca_objective}) can be considered as a negative joint log-posterior for $\vec{p}$ and $\vec{c}_V$, where the first
quadratic form represents the log-likelihood, and the regularization
term can be viewed as a log-prior density for the source power
cross-spectra.  

\section{Spectral model for AME} \label{sec:appa}
Theoretical spinning dust models predict a variety of spectra, which can be substantially different in shape, depending on a large number of parameters describing the physics of the medium. The number of such physical parameters is too large to be constrained by the data in the available frequency range. For the purpose of the estimation of the spectral behaviour of the AME we adopt a simple formula depending on only a few parameters. The CCA component separation method used in this work implements the parametric relation proposed by \cite{bonaldi2007} [Eq.~(\ref{tspin})], depending on the peak frequency, $\nu_{\rm p}$, and slope at 60\,GHz, $m_{60}$. To verify the adequacy of this parametrization we produced spinning dust spectra for different input physical parameters with the {\tt SpDust} code and fitted each of them with the proposed relation by minimizing the $\chi^2$ for the set of frequencies used in this work. The input models we consider are: weak neutral medium (WNM); cold neutral medium (CNM); weak ionized medium (WIM); and molecular cloud (MC). Both the input {\tt SpDust} parameters and the best-fit $m_{60}$, $\nu_{\rm p}$ parameters for each model are reported in Table~\ref{tab:commontab}. 
For comparison, we also consider alternative parametric relations and in particular:
\begin{itemize}
\item the model implemented in the {\tt Commander} component separation method (\citealt{pietrobon}, \citealt{planck_haze}) which is a Gaussian in the $T_{\rm CMB}-\ln(\nu)$ plane, parametrized in terms of central frequency and width;
\item the \cite{tegmark2000} model, which is a modified black-body relation [Eq.~(\ref{dust})] having  temperature around $0.25\,$K and emissivity index around 2.4;
\end{itemize}
As this test does not account for the presence of the other components and does not include any data simulation, it verifies the intrinsic ability of the parametric model to reproduce the actual spectra. Realistic estimation errors for the CCA model are derived through simulations in Appendix~\ref{sec:simulations}.

Fig.~\ref{fig:ametest} compares the input spectra with the best-fit models for the different parametrizations. In general, the fits are accurate at least up to  $\nu = 50$--60\,GHz, while at higher frequencies the parametric relations may not be able to reproduce the input spectra in detail.  This is a consequence of fitting complex spectra with only a few parameters. The fit tends to fail where the AME signal is weaker. 

Over the frequency range considered, CCA and {\tt Commander} models fit the input spectrum generally better than the \cite{tegmark2000} model (which falls off too rapidly at high frequencies). When adding lower frequency data, however, CCA and {\tt Commander} models will be increasingly inaccurate, as they are symmetric with respect to the peak of the emission.  
The models implemented by CCA and {\tt Commander} perform quite similarly, despite the different formulation. As a result, those methods are able to give consistent answers, which ensures consistency between different analyses within \Planck~\citep[e.g.][]{planck_haze}.

The CCA model used in this work provides a reasonable fit to theoretical spinning dust models for a variety of physical conditions. 
The best-fit parameters that we obtain, reported in Table~\ref{tab:commontab}, vary significantly from one input model to another and have a straightforward interpretation in terms of the spectrum. 

\begin{table*}[tmb]                 % table* is a two-column table.  Drop the * for one column.
\begingroup
\newdimen\tblskip \tblskip=5pt
%\caption{}                          % Caption goes here.
\caption{{\tt SpDust} input parameters for the spectra in Fig.~\ref{fig:ametest} and best-fit parameters for the CCA spectral model. The {\tt SpDust} input parameters are: the total hydrogen number density $n_{\rm H}$, the gas temperature $T$, the intensity of the radiation field relative to the average interstellar radiation field $\chi$, the hydrogen ionization fraction $x_{\rm H}=n_{\rm H^{+}} / n_{\rm H}$ and the ionized carbon fractional abundance $x_{\rm C}=n_{\rm C^{+}} / n_{\rm H}$.}
%\label{}                            % Label goes here.
\label{tab:commontab}
\nointerlineskip
\vskip -3mm
\footnotesize
\setbox\tablebox=\vbox{
   \newdimen\digitwidth 
   \setbox0=\hbox{\rm 0} 
   \digitwidth=\wd0 
   \catcode`*=\active 
   \def*{\kern\digitwidth}
   \newdimen\signwidth 
   \setbox0=\hbox{+} 
   \signwidth=\wd0 
   \catcode`!=\active 
   \def!{\kern\signwidth}
{\tabskip=2em
\halign{\hfil#\hfil & #\hfil& #\hfil & #\hfil & #\hfil & #\hfil & #\hfil & #\hfil &#\hfil&#\hfil \cr                          % Template goes here.
\noalign{\doubleline}
&{\tt SpDust}&&&&&&CCA&\cr
model name&nH [cm-3]&T[K]&$\chi$&$x_{\rm H}$&$x_{\rm C}$&&$\nu_{\rm p}$&$m_{60}$\cr
\noalign{\vskip 3pt\hrule\vskip 5pt}
WNM &***0.4&6000 &1.00 &0.10  &0.0003 & &24.22&7.53  \cr
CNM &**30.0&*100 &1.00 &0.0012&0.0003 & &29.00&4.93  \cr
WIM &***0.1&8000 &1.00 &0.99  &0.001  & &27.30&5.66  \cr
MC  &*300  &**20 &0.01 &0.0   &0.0001 & &38.77&2.00  \cr
\noalign{\vskip 5pt\hrule\vskip 3pt}}}
}
%\endPlancktable                    % ends one-column \halign
\endPlancktablewide                 % ends two-column \halign
%\tablenote a Footnote a.\par
%\tablenote b Footnote b.\par
\endgroup
\end{table*}                        % table* is a two-column table.  Drop the * for one column.
\begin{figure}
%figb1
\begin{center}
\includegraphics[width=88mm]{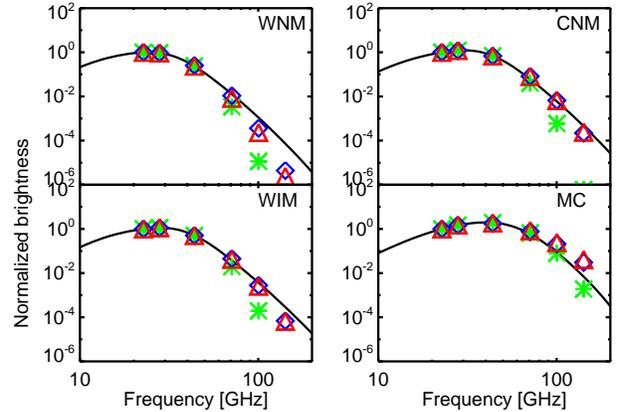}
\caption{Theoretical spinning dust models produced with {\tt SpDust} (solid lines) and fitted with CCA (triangles), {\tt Commander} (diamonds), and \cite{tegmark2000} (asterisks) models. Input {\tt SpDust} parameters and best-fit parameters for the CCA model are provided in Table~\ref{tab:commontab}.}
\label{fig:ametest}
\end{center}
\end{figure}
\section{Description of the simulations}\label{sec:simulations}
\begin{figure*}
%figc1
\begin{center}
\begin{tabular}{cccc}
\includegraphics[width=4.25cm]{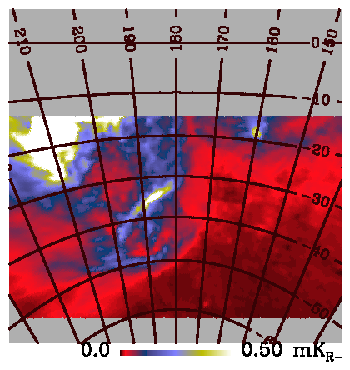}&\includegraphics[width=4.25cm]{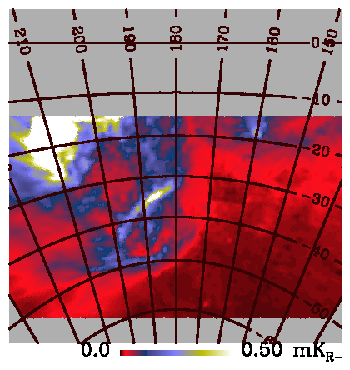}&\includegraphics[width=4.25cm]{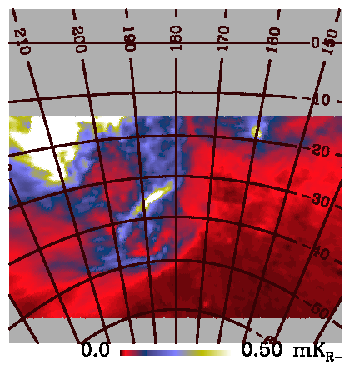}&\includegraphics[width=4.25cm]{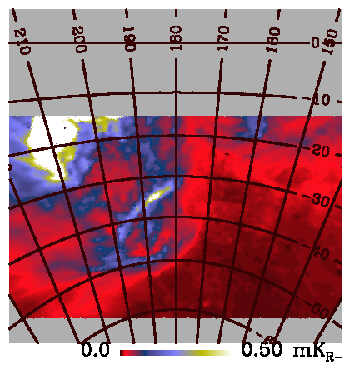}\\
&\includegraphics[width=4.25cm]{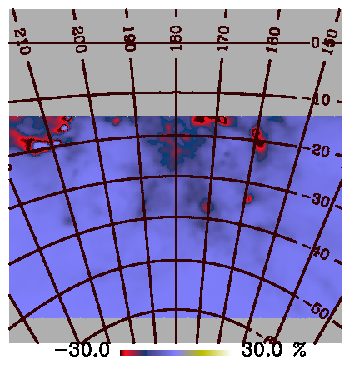}&\includegraphics[width=4.25cm]{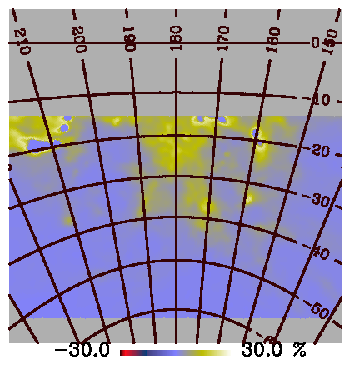}&\includegraphics[width=4.25cm]{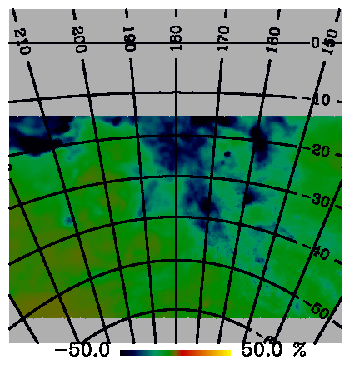}\\
\end{tabular}
\caption{Free-free templates at 23\,GHz used for the analysis. The reference template FF$_{\rm{REF}}$ is in the upper left corner; the other columns ({\it left to right}) are  FF$_{\rm{1}}$,  FF$_{\rm{2}}$ and FF$_{\rm{3}}$, respectively. The differences in the lower panels $(\rm{FF}_{\rm{i}}-\rm{FF}_{\rm{REF}})/\rm{FF}_{\rm{REF}}$ are on average of the order of 10\,\%, but reach 50\,\% in regions of strong dust emission.}
\label{fig:ff_template}
\end{center}
\end{figure*}
We simulated \Planck~and {\it WMAP} 7-yr data by assuming monochromatic bandpasses positioned at the central frequency of the bands, Gaussian beams at the nominal values indicated in Tables \ref{tab:planckdata} and \ref{tab:ancillarydata}, and Gaussian noise generated according to realistic, spatially varying noise RMS. Our model of the sky consists of the following components:
\begin{itemize}
\item CMB emission given by the best-fit power spectrum model from {\it WMAP} 7-yr analyses;
\item synchrotron emission given by  the \cite{haslam} template scaled in frequency with a power-law model with a spatially varying synchrotron spectral index $\beta_{\rm s}$, as modelled by \cite{giardino};
\item free-free emission given by the \cite{cliveff} H$\alpha$ corrected for dust absorption with the $E(B-V)$ map from \cite{schlegel} with a dust absorption fraction  $f_{\rm d}=0.33$, and scaled in frequency according to Eq.~(\ref{ff}) with $T_{\rm e}=7000$\,K;
\item thermal dust emission modelled with the 100$\microns$ map from \cite{schlegel}, scaled in frequency according to Eq.~(\ref{dust}) with $T_{\rm d}=18$\,K and a spatially varying $\beta_{\rm d}$ having average value of 1.7;
\item AME modelled by the $E(B-V)$ map from \cite{schlegel} with intensity at 23\,GHz calibrated using the results of \cite{ghosh2011} for the same region of the sky. 
\end{itemize} 

We adopted more than one spectral model for the AME. We first considered two convex spectra, generated with the {\tt SpDust} code: one peaking around  26\,GHz and the other  peaking around 19\,GHz. We also tested a spatially varying power-law model (with spectral index of $-3.6 \pm 0.6$, \citealt{ghosh2011}), which could result from the superposition of multiple convex components along the line of sight.

It is worth noting that the simulated sky is more complex than the model assumed in the component separation. This has been done intentionally, to reflect a more realistic situation. Another realistic feature we included is the presence of errors in the synchrotron and free-free templates. 
The spatial variability of the synchrotron spectral index modifies the morphology of the component with respect to that traced by the 408\,MHz map from \cite{haslam}. 
The use of H$\alpha$ as a tracer of free-free emission is affected by even larger uncertainties. Our uncertainties on the dust absorption fraction $f_{\rm d}$ (estimated to be $f_{\rm d}=0.33^{+0.10}_{-0.15}$ at intermediate latitudes by \citealt{cliveff}) and on the scattering of H$\alpha$ photons from dust grains, can create dust-correlated biases in the template. This is illustrated in Fig.~\ref{fig:ff_template}, where we compare different versions of the free-free template. FF$_{\rm REF}$ is our reference template, adopted for the analysis of real data and for simulating the component, which is corrected for $f_{\rm d}=0.33$ as described in \cite{cliveff}. Two more templates (FF$_{\rm{1}}$ and FF$_{\rm{2}}$) have been obtained by correcting H$\alpha$ for $f_{\rm d}=0.33-0.15$ and $f_{\rm d}=0.33+0.1$ ($\pm 1\sigma$ according to \citealt{cliveff}). A final template (FF$_{\rm{3}}$) has been obtained by correcting FF$_{\rm{REF}}$ for scattered light at the 15\,\% level by subtracting from the free-free map the 1$\microns$ map of \cite{schlegel} multiplied by a suitable constant factor \citep{witt}. Difference maps  $(\rm{FF}_{\rm{i}}-\rm{FF}_{\rm{REF}})/\rm{FF}_{\rm{REF}}$, presented in the lower panels of Fig.~\ref{fig:ff_template}, are on average of order of 10\,\%, but can be much higher (up to 50--60\,\%) in regions of strong dust emission. 

\begin{figure}
%figC2
\begin{center}
\begin{tabular}{cc}
\includegraphics[width=4cm]{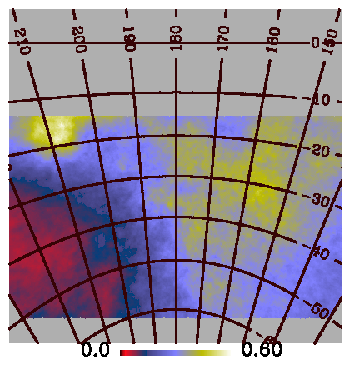}&\includegraphics[width=4cm]{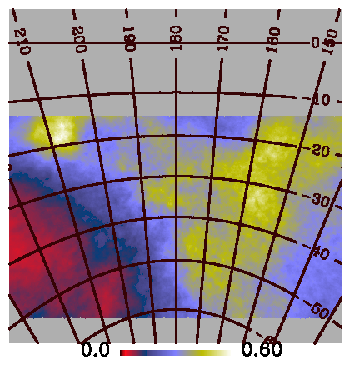}\\
&\includegraphics[width=4cm]{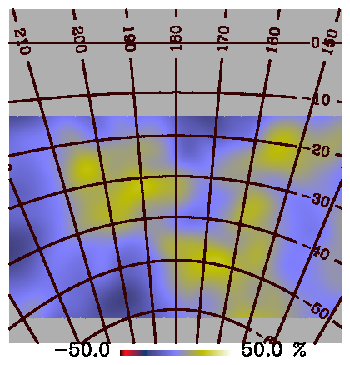}\\
\end{tabular}
\caption{{\it Upper panel}: simulated synchrotron component ({\it left}) and synchrotron template ({\it right}) at 23\,GHz. {\it Lower panel}: difference map divided by the simulated component.}
\label{fig:syn_template}
\end{center}
\end{figure}

\begin{figure}
%fig5
\begin{center}
\includegraphics[width=8.8cm]{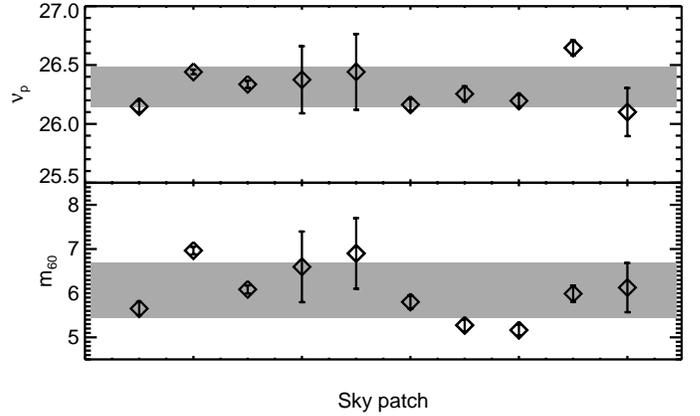}
\caption{Average and RMS of $\nu_{\rm p}$ and $m_{60}$ estimated over simulations having different CMB and noise realizations for different patches on the $x$-axis. The grey area is the average and RMS over different patches, which is typically larger than that due to noise and CMB.}
\label{fig:CMB_sim}
\end{center}
\end{figure}

When analysing the simulated data, we used both FF$_{1}$ and FF$_{2}$ as free-free templates in place of FF$_{\rm REF}$, which corresponds to the simulated component. For synchrotron emission the morphological mismatch between the simulated component and the template has been achieved by scaling  the component  from 23\,GHz to 408\,MHz with a spatially varying spectral index. The comparison between component and template is presented in Fig.~\ref{fig:syn_template}; the differences are of the order of 10\,\%. 
The simulated data-sets described above have been analysed with the CCA method using the same procedure applied to the real data; the results of this assessment are presented in Sect.~\ref{sec:simul_0}. 
As a separate test, we verified the impact of the CMB component on the results for $\nu_{\rm p}$  and $m_{60}$. We generated 100 sets of mock data having the same foreground emission and different realizations of CMB and instrumental noise, and repeated the estimation of the AME frequency scaling. For this test we used the simulation with a spatially constant AME spectrum peaking at 26\,GHz. As this analysis is computationally demanding, the CCA estimation has been performed only on the 10 independent patches covering the Gould Belt region (centred on latitudes $-20^{\circ}$ and $-40^{\circ}$ and longitudes of $140^{\circ}$, $160^{\circ}$, $180^{\circ}$, $200^{\circ}$, and $220^{\circ}$). In Fig. \ref{fig:CMB_sim} we show the average (diamonds) and  RMS (error bars) $\nu_{\rm p}$ and $m_{60}$ over the 100 realizations for each patch, for different patches on the $x$-axis. The scatter between the results obtained for different patches (indicated by the grey area in the plots) is typically larger than the error bars, measuring the scatter due to different CMB realizations. This means that the foreground emission generally dominates over the CMB as a source of error. Larger error bars associated with the CMB are obtained for three patches having fainter foreground emission. For such patches the estimated errors on $\nu_{\rm p}$ and $m_{60}$ are consistently larger. The CMB variation results on average in $\Delta \nu_{\rm p}=0.1$\,GHz and $\Delta m_{60}=0.3$, which reach 0.3\,GHz and 0.8, respectively, for the worst sky patch. Those values are below the error bars resulting from the analysis of the data, which amount to 1--1.5\,GHz for $\nu_{\rm p}$ and 1.5--2 for $m_{60}$. The CMB  has limited impact on the results because this component is modelled in the mixing matrix. Having a known frequency scaling, the statistical constraint used by CCA is able to trace the pattern of the CMB through the frequencies with good precision and hence identify it correctly.

%\newpage
\raggedright
\end{document}